\documentclass{article}

\usepackage[final]{neurips_data_2024}

\usepackage[utf8]{inputenc} % allow utf-8 input
\usepackage[T1]{fontenc}    % use 8-bit T1 fonts
\usepackage{hyperref}       % hyperlinks
\usepackage{url}            % simple URL typesetting
\usepackage{booktabs}       % professional-quality tables
\usepackage{amsfonts}       % blackboard math symbols
\usepackage{nicefrac}       % compact symbols for 1/2, etc.
\usepackage{microtype}      % microtypography
\usepackage{xcolor}         % colors
\usepackage{multirow}
\usepackage{graphicx}
\usepackage{arydshln}
\usepackage{xspace}
\usepackage{listings}
\usepackage{stix}
\usepackage{fontawesome5}
\usepackage{subcaption} 
\usepackage{pifont}
\usepackage{tablefootnote}
\usepackage{threeparttable}
\usepackage{url}

\newcommand{\sys}{\textsc{ConGra}\xspace}
\newcommand{\va}{$A$\xspace}
\newcommand{\vb}{$B$\xspace}
\newcommand{\base}{$O$\xspace}
\newcommand{\merged}{$M$\xspace}
\newcommand{\resolved}{$R$\xspace}

\newcommand{\lstbg}[3][0pt]{{\fboxsep#1\colorbox{#2}{\strut #3}}}
\definecolor{codegreen}{rgb}{0,0.6,0}
\lstdefinelanguage{diff}{
	frame=shadowbox,
	basicstyle=\ttfamily\scriptsize\bfseries,
        breaklines=true,
	morecomment=[f][\color{red}]{---}, 
	morecomment=[f][\color{codegreen}]{+++},
        morecomment=[f][\color{orange}]{<<<<<<<},
        morecomment=[f][\color{orange}]{=======},
        morecomment=[f][\color{orange}]{>>>>>>>},
	morecomment=[f][\lstbg{red!20}]{-\ },
	morecomment=[f][\lstbg{green!20}]{+\ },
	morecomment=[f][\color{blue}]{@@},
}

\usepackage{tcolorbox}
\newcounter{findingCounter}
\title{\sys: Benchmarking Automatic Conflict Resolution}
\author{%
  \textbf{Qingyu~Zhang, Liangcai~Su, Kai~Ye, Chenxiong~Qian} \\ The University of Hong Kong\\
}

\begin{document}

\maketitle

\begin{abstract}
Resolving conflicts from merging different software versions is a challenging task. 
%
%Various tools have been developed to reduce the cost of manual merges. 
%To reduce the cost of manual merges, various tools have been developed. 
To reduce the overhead of manual merging, researchers develop various program analysis-based tools which only solve specific types of conflicts and have a limited scope of application.
%
%With advancements in large language models (LLMs), researchers are leveraging their language-awareness capabilities for conflict resolution. 
With the development of language models, researchers treat conflict code as text, which theoretically allows for addressing almost all types of conflicts.
However, the absence of effective conflict difficulty grading methods hinders a comprehensive evaluation of large language models (LLMs), making it difficult to gain a deeper understanding of their limitations. 
Furthermore, there is a notable lack of large-scale open benchmarks for evaluating the performance of LLMs in automatic conflict resolution.
%
% 现有的ACR工具只能克服的冲突范围是很有限的。
% 为此，近期研究人员将研究重心放到了大语言模型上，将冲突视为文本，理论上能够适用于所有的冲突。
% 然而，\textbf{对于大语言模型的冲突解决能力评估并不容易}。 
% 尽管冲突的数据很丰富，但目前缺乏有效的冲突难度分级方式，无法充分评估LLM的各方面能力。 
% 此外，现有方法缺乏对于大语言模型的综合评估，尤其是针对于冲突上下文文本十分丰富的情况下。
% 
%However, proper evaluation of code merging techniques remains elusive, as the community struggles to converge on methodologies and standard metrics, often overlooking the performance of merge tools in resolving conflicts of varying difficulties.
%
To address these issues, we introduce \sys, a \textbf{CON}flict-\textbf{GRA}ded benchmarking scheme designed to evaluate the performance of software merging tools under varying complexity conflict scenarios.
We propose a novel approach to classify conflicts based on code operations and use it to build a large-scale evaluation dataset based on 44,948 conflicts from 34 real-world projects. 
We evaluate state-of-the-art LLMs on conflict resolution tasks using this dataset. 
By employing the dataset, we assess the performance of multiple state-of-the-art LLMs and code LLMs, ultimately uncovering two counterintuitive yet insightful phenomena.
% Our evaluation reveals two counterintuitive yet highly significant findings that will enhance the understanding of applying LLMs to conflict resolution tasks. 
%
\sys will be released at \href{https://github.com/HKU-System-Security-Lab/ConGra}{https://github.com/HKU-System-Security-Lab/ConGra}.
\end{abstract}

\section{Introduction}

% outline
% 1. Development of traditional merging tools
% 2. Development of machine learning merging tools
% 3. Current evaluation limitation
% 4. Our approach & eval & conclusion

%
Code merging has become a challenging task for developers during project development and maintenance. 
Git, the most popular version control system (\cite{spinellis2012git}), uses a text-based code-merging mechanism. Despite its efficiency, developers often struggle with manually merging different versions of code when Git fails to resolve conflicts automatically. 
These conflicts come from text or even the syntax and the functionality of the code from different versions.

To address conflict resolution, researchers leverage program analysis to achieve syntax-error-free code merging (\cite{mastery, safemerge, spork, intellimerge, jdime, fstmerge}). 
Based on abstract syntax trees (AST), these tools merge AST vertices and edges to ensure syntax correctness and better merge results. 
Nevertheless, a conflict is still generated and need to be resolved by developers if the merge of AST nodes fails. 
With the development of language models and even LLMs, they are now applied to conflict resolution (\cite{deepmerge, gmerge, MergeBERT, mergegen}). 
Trained on extensive datasets of prior and manually merged codes, these models predict suitable resolution for each conflicting code segment without manual efforts.

However, evaluating the performance of LLMs on conflict resolution tasks is challenging due to the wide variation in conflict difficulty and the lack of effective grading methods to reflect these differences. 
For example, ConflictBench (\cite{shen2024conflictbench}) classifies conflicts based on the source of conflict-resolved code, such as from either merging candidate versions or newly introduced by developers.
This classification does not accurately reflect the complexity of the conflicts.
Additionally, the community lacks comprehensive conflict resolution benchmarks (discussed in \S \ref{sc:conflict-res-benchmark}), especially for extreme cases involving long code contexts.

To this end, we introduce \sys, which is designed to evaluate code merging tools across a diverse range of merging scenarios and assess their ability to resolve conflicts of varying complexities.
We propose a novel approach to construct graded conflict dataset according to conflict's resolving complexity.
We evaluated six state-of-the-art LLMs (three general LLMs and three code LLMs) on \sys to assess their abilities to resolve conflict in various merging scenarios.
The dataset is constructed using 44,948 conflict cases sourced from 34 large-scale open-source projects written in C, C++, Java, and Python. 
The results show that LLMs with longer contexts support do not always yield better results compared to models with shorter contexts.
Additionally, general LLMs (e.g. LLama3-8B and DeepSeek-V2) outperform specialized code LLMs in automatically resolving conflict.
Besides, we will release our datasets and a benchmark at \href{https://github.com/HKU-System-Security-Lab/ConGra}{https://github.com/HKU-System-Security-Lab/ConGra}.

In summary, we made the following contributions:

\begin{enumerate}
    \item We introduce the first classification approach to generate a complexity-graded conflict dataset. 
    Using this appraoch, collected conflicts can be classified into seven categories.

    \item We release a large-scale graded dataset for conflict resolution benchmarking. which contains 44,948 conflict cases from popular projects written in C, C++, Java, and Python.

    \item We conduct the first comprehensive evaluation of LLM's performance on conflict resolution task, and find two thought-provoking counter-intuitive phenomena.
    
\end{enumerate}
\section{Background}
\subsection{Task Definition of Automatic Merge Conflict Resolution}

Git offers multiple strategies to merge code (\cite{gitmergealgorithm}). 
The most common and widely-used strategy is the three-way merging strategy (\cite{three-way-merge}).
The fundamental concept behind three-way merging is to locate the most recent common ancestor (\base) through the historical commit graph, given two merging candidates (\va and \vb), and to generate the automerged code version (\merged) based on the difference $A-O$ and the difference $B-O$.
Before generating \merged, Git identifies multiple difference matching triplets $A_i-O_i-B_i$ by comparing $A-O$ and $B-O$. 
For each difference matching triplet, 
1) if $A_i-O_i$ equals $B_i-O_i$, it indicates that both \va and \vb have made identical changes to the same code segment. In this case, the difference block will be applied to \merged; 
2) if $A_i-O_i$ is empty, but $B_i-O_i$ exists, it suggests that only \vb has modified this code segment. Consequently, the difference block from \vb will be applied to \merged, and vice versa;
3) if $A_i-O_i$ and $B_i-O_i$ are not empty nor equal, it indicates that \va and \vb have made different modifications to the same code segment, resulting in a \textit{conflict} in \merged. In this paper, we assume that \merged always contains at least one conflict.

Due to the presence of conflicts, Git-based automated code merging may encounter exceptions, prompting developers to manually address them. Following conflict resolution, developers will release the resolved code version (\resolved).
In large projects, merging two versions results in numerous conflicts, involving substantial code modifications. This significantly raises developers' project maintenance costs (\cite{vale2021challenges}).
To this end, various automatic conflict resolution (ACR) systems have been proposed to mitigate these challenges.

\begin{table}[ht]
\centering
\caption{Evaluation features implemented in 1) program analysis-based ACR (tagged with \dag); 2) machine learning-based ACR (tagged with \ddag); 3) conflict resolution benchmark (tagged with \maltese). The symbol "\textbf{-}" means the feature is not applicable to the corresponding work's evaluation.}
\label{tb:afeature_compare}
% \resizebox{\textwidth}{!}{
\begin{threeparttable}
\begin{tabular}{@{}lllllll@{}}
\toprule
Evaluation Features  & \multicolumn{1}{c}{$Feature_1$} & \multicolumn{1}{c}{$Feature_2$} & \multicolumn{1}{c}{$Feature_3$} & \multicolumn{1}{c}{$Feature_4$} & \multicolumn{1}{c}{$Feature_5$} & \multicolumn{1}{c}{$Feature_6$} \\ \midrule
IntelliMerge\dag        & \textcolor{green}{\ding{51}}                            & \textcolor{red}{\ding{55}}                            & \textcolor{green}{\ding{51}}                            & \textcolor{green}{\ding{51}}                            & \textcolor{green}{\ding{51}}                            & \textcolor{red}{\ding{55}}                            \\
Spork\dag               & \textcolor{red}{\ding{55}}                            & \textcolor{red}{\ding{55}}                            & \textcolor{green}{\ding{51}}                            & \textcolor{green}{\ding{51}}                            & \textcolor{green}{\ding{51}}                            & \textcolor{red}{\ding{55}}                            \\
SafeMerge\dag           & \textcolor{red}{\ding{55}}                            & \textcolor{red}{\ding{55}}                            & \textcolor{green}{\ding{51}}                            & \textcolor{red}{\ding{55}}                            & \textcolor{red}{\ding{55}}                            & \textcolor{red}{\ding{55}}                            \\
FSTMerge\dag            & \textcolor{green}{\ding{51}}                            & \textcolor{red}{\ding{55}}                            & \textcolor{green}{\ding{51}}                            & \textcolor{red}{\ding{55}}                            & \textcolor{red}{\ding{55}}                            & \textcolor{red}{\ding{55}}                            \\
Mastery\dag             & \textcolor{red}{\ding{55}}                            & \textcolor{red}{\ding{55}}                            & \textcolor{green}{\ding{51}}                            & \textcolor{red}{\ding{55}}                            & \textcolor{green}{\ding{51}}                            & \textcolor{red}{\ding{55}}                            \\
JDIME\dag               & \textcolor{green}{\ding{51}}                            & \textcolor{red}{\ding{55}}                            & \textcolor{green}{\ding{51}}                            & \textcolor{red}{\ding{55}}                            & \textcolor{red}{\ding{55}}                            & \textcolor{red}{\ding{55}}                            \\
MergeBERT\ddag           & \textcolor{red}{\ding{55}}                            & \textcolor{red}{\ding{55}}                            & \textbf{-}                            & \textcolor{green}{\ding{51}}                            & \textcolor{green}{\ding{51}}                            & \textcolor{red}{\ding{55}}                            \\
GMerge\ddag              & \textcolor{red}{\ding{55}}                            & \textcolor{red}{\ding{55}}                            & \textbf{-}                            & \textcolor{green}{\ding{51}}                            & \textcolor{green}{\ding{51}}                            & \textcolor{red}{\ding{55}}                            \\
DeepMerge\ddag           & \textcolor{red}{\ding{55}}                            & \textcolor{red}{\ding{55}}                            & \textbf{-}                            & \textcolor{green}{\ding{51}}                            & \textcolor{green}{\ding{51}}                            & \textcolor{red}{\ding{55}}                            \\
MergeGen\ddag            & \textcolor{red}{\ding{55}}                            & \textcolor{red}{\ding{55}}                            & \textbf{-}                            & \textcolor{green}{\ding{51}}                            & \textcolor{green}{\ding{51}}                            & \textcolor{red}{\ding{55}}                            \\
ConflictBench\maltese
       & \textcolor{green}{\ding{51}}                            & \textcolor{red}{\ding{55}}                            & \textcolor{green}{\ding{51}}                            & \textcolor{green}{\ding{51}}                            & \textcolor{green}{\ding{51}}                            & \textcolor{red}{\ding{55}}                            \\
\sys\maltese
 & \textcolor{green}{\ding{51}}                            & \textcolor{green}{\ding{51}}                            & \textcolor{green}{\ding{51}}                            & \textcolor{green}{\ding{51}}                            & \textcolor{green}{\ding{51}}                            & \textcolor{green}{\ding{51}}                            \\ \bottomrule
\end{tabular}
% }

\begin{tablenotes}
\footnotesize
\item \textbf{$Feature_1$: }Classify dataset; \textbf{$Feature_2$: }Assess performance under different graded conflicts; \textbf{$Feature_3$: }Record number of generated conflicts; \textbf{$Feature_4$: }Calculate precision of the generated conflict resolution; \textbf{$Feature_5$: }Calculate accuracy of the generated conflict resolution; \textbf{$Feature_6$: }Assess the whole code similarity instead of using exact string matching.
\end{tablenotes}
\end{threeparttable}

\end{table}

\subsection{Program Analysis-based ACR}
\label{sc:program-ana-acr}

% Program analysis-based ACR is divided into structural and semi-structural merging tools. 
% %
% Structural merging tools (\cite{mastery, spork, safemerge}) convert the code (\va, \vb, \base) into an abstract syntax tree (AST) and compare corresponding nodes on the AST, adhering to the core principle of the original three-way merging. 
% %
% Semi-structural merging (\cite{fstmerge, intellimerge, jdime}) combines structural and textual merging. 
% %
% Structural merging is used for syntax-sensitive code blocks, while textual merging is applied to syntax-insensitive content like comments and strings (\cite{accioly2018understanding}). 
%
Program analysis-based ACR (\cite{mastery, spork, safemerge, fstmerge, intellimerge, jdime}) ensures the syntactic integrity of the merged code by merging on AST.
As they implement the three-way merging strategy, conflicts inevitably arise for unmergeable AST nodes, requiring manual intervention.

To assess the performance of program analysis-based ACR, researchers focus on metrics like the number of generated conflicts, the accuracy of resolutions, and resource consumption. 
However, these tools often lack convincing evaluation datasets and performance comparisons.
As shown in \autoref{tb:afeature_compare}, tools like Spork, SafeMerge, and Mastery (\cite{spork, safemerge, mastery}) evaluate all merging scenarios without classification, obscuring performance differences across conflict complexities. 
JDIME and FSTMerge (\cite{jdime, fstmerge}) separately categorize structural and textual merges, but this is essentially an ablation study, leaving the dataset's resolving complexity intertwined.
IntelliMerge (\cite{intellimerge}) classifies merge scenarios but fails to differentiate these categories in the final evaluation, preventing performance comparisons across scenarios.
Additionally, These tools evaluate the quality of generated resolutions by exact string matching with the manual resolution (ground truth). 
However, this approach is inadequate for measuring code similarity, as it is influenced by programming style and code structure.

\subsection{Machine learning-based ACR}
Machine learning-based ACR (\cite{gmerge, mergegen, MergeBERT, deepmerge}) leverages a vast number of code merge examples during pre-training and harnesses the language-aware capabilities of machine learning models to understand the intricacies of code merging. 
Once trained, these models offer resolution suggestions for merge conflicts. 
They does not rely on the three-way merge strategy, eliminating the need for manual conflict resolution.

~\autoref{tb:afeature_compare} illustrates the features implemented in machine learning-based ACRs' evaluation.
Machine learning-based ACRs must provide accurate resolutions for all conflicts, underscoring the importance of resolution correctness.
Prior research efforts have primarily concentrated on improving two key metrics: precision and accuracy. Accuracy measures the percentage of total conflicts for which these tools produce the correct resolution, while precision indicates the percentage of correct resolutions among all resolution suggestions provided by the tools.
By the same token as depicted in \S \ref{sc:program-ana-acr}, the matching algorithm between generated resolution and ground truth are exact string matching, which cannot reflect the real code similarity.
Furthermore, existing language model-based ACR test datasets also lack classification of merge scenarios based on different complexities, hindering the assessment of model performance across various levels of code merging tasks' difficulty.

\subsection{Conflict Resolution Benchmarks}
\label{sc:conflict-res-benchmark}
% Complexity-aware 
% *Large* Language Models Results 
% Abundant Indicators

%
ConflictBench (\cite{shen2024conflictbench}) is the only benchmark for software code merging evaluation to date. 
It categorizes merging scenarios by resolution types and uses three metrics with an exact string matching algorithm to evaluate merging tools. 
However, as shown in \autoref{tb:afeature_compare}, its classification strategy doesn't fully capture the complexity of merging scenarios.
For example, conflicts can be resolved by retaining all edits from \va or \vb, or by introducing new edits. 
ConflictBench categorizes these into three groups, but the complexity boundaries between them are often blurred. 
Additionally, ConflictBench's datasets include only 180 merging cases, and further classification limits the data volume in each category, potentially leading to inaccurate evaluation results.

% Related Work 
% https://github.com/ICSE-2022-Submission/Automatic-Merge-Conflict-Resolution-Tools/tree/main
\section{Benchmarking Pipeline}
\label{sec:pipeline}
\subsection{Overview}

\begin{figure}[th]
	\centerline{\includegraphics[width=\linewidth]{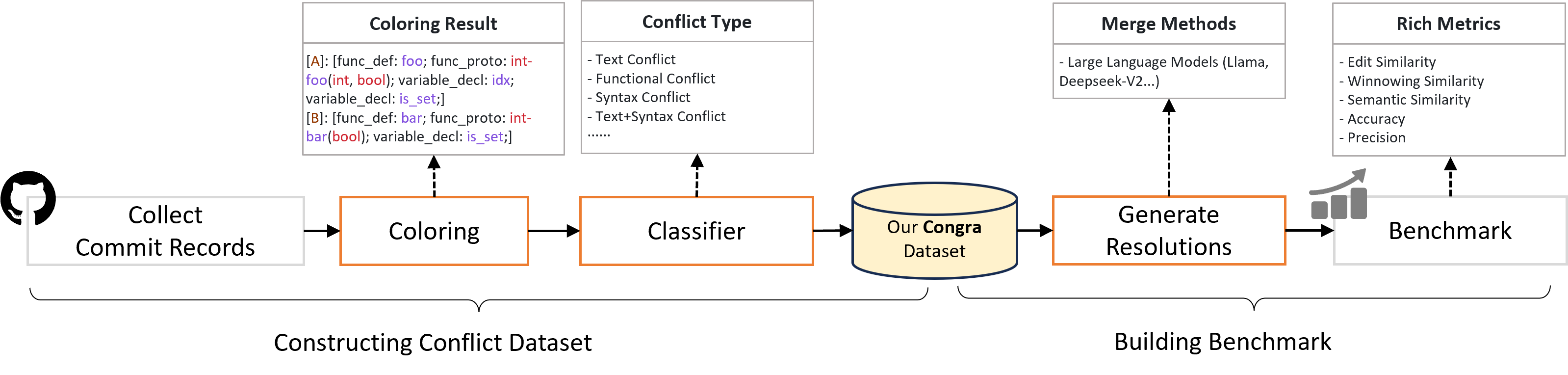}}
	\caption{\textmd{Benchmarking pipeline overview.}}
	\label{fg:pipline_overview}
\end{figure}

\autoref{fg:pipline_overview} provides an overview of \sys's benchmarking pipeline.
\sys starts with collecting open-source projects' historical merging scenarios with conflict from Github.
After the raw dataset is constructed, \sys further colors the conflict snippets with code operations which are extracted via a lightweight but powerful syntax tree level analysis.
These conflicts then are classified into various categories based on the code operations taken in the conflict code block to construct \sys dataset.
Then, we collect the conflict along with its context to construct the prompt for LLMs.
Finally, LLMs will respond to the query and provide conflict resolution suggestions.
The results will be assessed by \sys's evaluation metrics.

\subsection{Coloring}

%
% Coloring stage is designed to deal with the following two tough tasks.

Firstly, analyzing conflict code snippets is challenging because the snippets in \merged generated by Git are often fragmented and can contain any part or format of code.
Existing program analysis tools (\cite{llvm, angr, soot}) cannot handle these incomplete snippets syntactically or semantically. 
To address this, \sys applies \merged's modifications from \va and \vb separately to generate two merged files ($M_a$ and $M_b$). 
This approach applies all the difference blocks from either \va or \vb, allowing for code analysis of the conflicting snippets by recording the text ranges and analyzing the whole code from $M_a$ and $M_b$.

Secondly, for the benchmarking pipeline, automating code complexity assessment must use the lightest code analysis. 
Traditional program analysis requires a strict program context, including the project environment and necessary compilation settings, which are hard to preserve in datasets. 
Additionally, program analysis tools perform data flow or control flow analysis (\cite{static-analysis, program-analysis}), consuming significant resources, especially for large-scale datasets.
\sys addresses this through code operation abstraction.
It extracts the operations of conflicting code fragments from $M_a$ and $M_b$ to reflect code complexity. 
Since \sys's analysis involves only lexical and grammatical analysis without deeper compilation or context dependency, it achieves efficient benchmarking.

Given $M_a$, $M_b$ and respective conflict-related code ranges, \sys extracts code operations from these conflict-related code blocks in the coloring phase.
\sys first leverages tree\_sitter (\cite{tree-sitter}), a lightweight multi-language code parsing framework, to convert plain text code into a syntax tree that is suitable for analysis.
We pre-define types of code operations based on the node of the syntax tree.
These code operations are arranged in ascending order of priority as shown below:

\begin{enumerate}
    \item \textbf{Composite Type Definition (CTD):} Definition of composite types, e.g. the definition of class and the definition of struct in C++ language.
    \item \textbf{Function Body Definition (FBD):} Definition of the function body content. Any modification to the function body is included in this operation.
    \item \textbf{Function Prototype Definition (FPD):} Definition of function prototype, i.e. function name, return value type and parameter list.
    \item \textbf{Language-Specific Operations (LSO):} Operations related to language features, such as macro definitions in C/C++ and the introduction of third-party libraries in Python.
    \item \textbf{Commenting (CMT):} Developers' comments on the code.
    \item \textbf{Variable Declaration (VD):} Declaration of all variables. These variables refers to global variables, local variables, and member variables defined in composite types.
\end{enumerate}

These code operations have different priorities, and operations with higher priorities can override operations with lower priorities.
For instance, the modification of a local variable name in a function body can belong to both function body definition and variable declaration. 
To this end, \sys will give priority to variable declaration as the result of code operation extraction. 
After analyzing each syntax tree node, \sys only retains code operations related to the conflicting code area, and generates code operation lists for the two versions of the code, namely $P_a$ and $P_b$.
These code operations serve as the basis for conflict classification in the subsequent stage.

\subsection{Classifier}
\label{sec:classifier}
\begin{figure}[!h]
    \centering
    \begin{subfigure}{0.32\textwidth}
        \includegraphics[width=\textwidth]{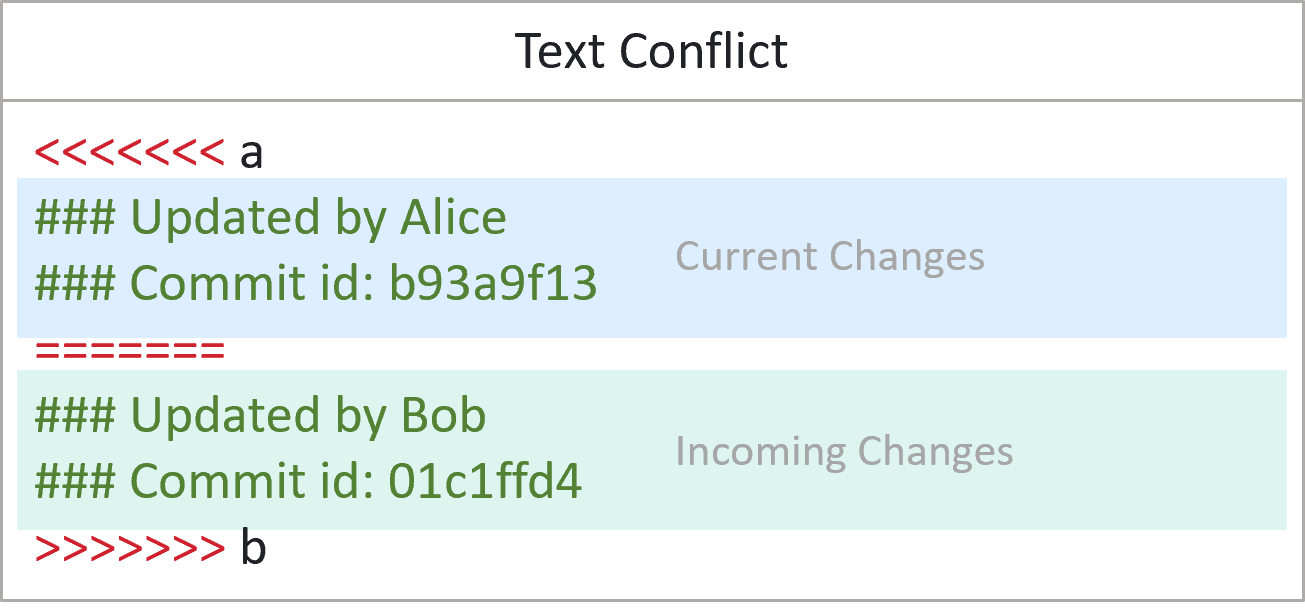}
        \caption{Text conflict.}
        \label{fig:text-conflict}
    \end{subfigure}
    \hfill
    \begin{subfigure}{0.32\textwidth}
        \includegraphics[width=\textwidth]{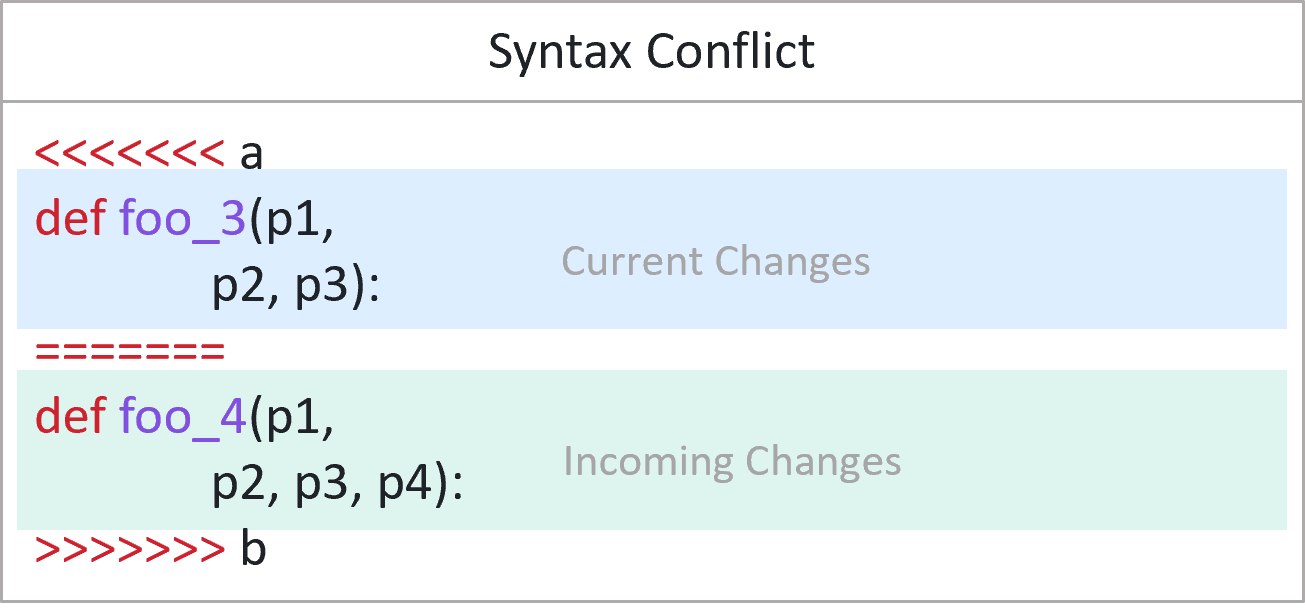}
        \caption{Syntax conflict.}
        \label{fig:syntax-conflict}
    \end{subfigure}
    \hfill
    \begin{subfigure}{0.32\textwidth}
        \includegraphics[width=\textwidth]{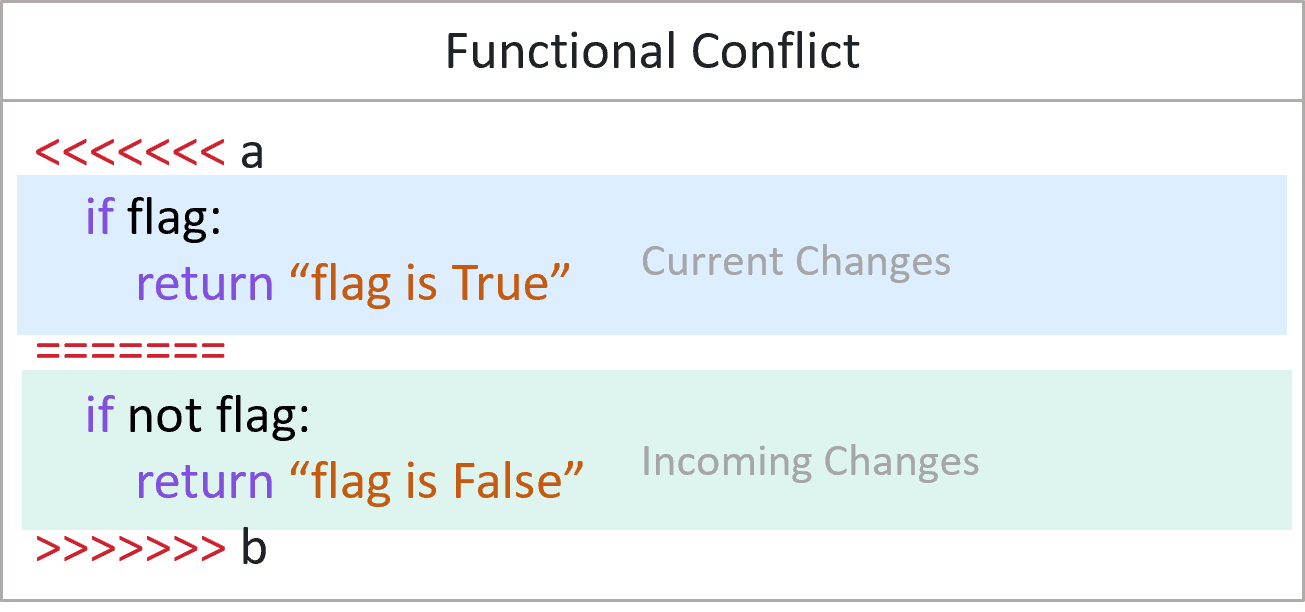}
        \caption{Functional conflict.}
        \label{fig:functional-conflict}
    \end{subfigure}
    \caption{Examples of text, syntax, and functional conflicts}
    \label{fig:conflict-type-example}
\end{figure}

Given $P_a$ and $P_b$, \sys traverses each pair of conflicting code blocks in \va and \vb and classifies the conflicts according to the following conditions. 
For each pair of \va’s conflict and \vb’s conflict (i.e., $P_{ai}$ in $P_a$ and $P_{bi}$ in $P_b$):

\begin{enumerate}
    \item \textbf{Text conflict: }At least one of $P_{ai}$ and $P_{bi}$ contains CMT. 
    ~\autoref{fig:text-conflict} presents a text conflict example.
    \item \textbf{Functional conflict: }At least one of $P_{ai}$ and $P_{bi}$ contains FBD.
    ~\autoref{fig:syntax-conflict} presents a functional conflict example.
    \item \textbf{Syntax conflict: }Respectively construct operations sets based on $P_{ai}$ and $P_{bi}$ (denoted as $S_a$ and $S_b$) with only remaining CTD, FPD, LSO, and VD. 
    Finally, the difference between $S_a$ and $S_b$ is not empty.
    ~\autoref{fig:functional-conflict} presents a syntax conflict example.
\end{enumerate}

Considering that the syntax among languages varies, the benchmarking of \sys only focuses on syntax problems caused by missing declarations or definitions which will occur in all programming languages. 
Since the classifier of \sys adopts the whitelist mode, the correctness of the classification results can be guaranteed.
As one conflict may be classified into multiple categories, \sys supports seven classifications, i.e., all permutations of text, syntax, and functional conflict.

\subsection{Generate Resolutions}
\begin{figure}[th]
\centerline{\includegraphics[width=0.6\linewidth]{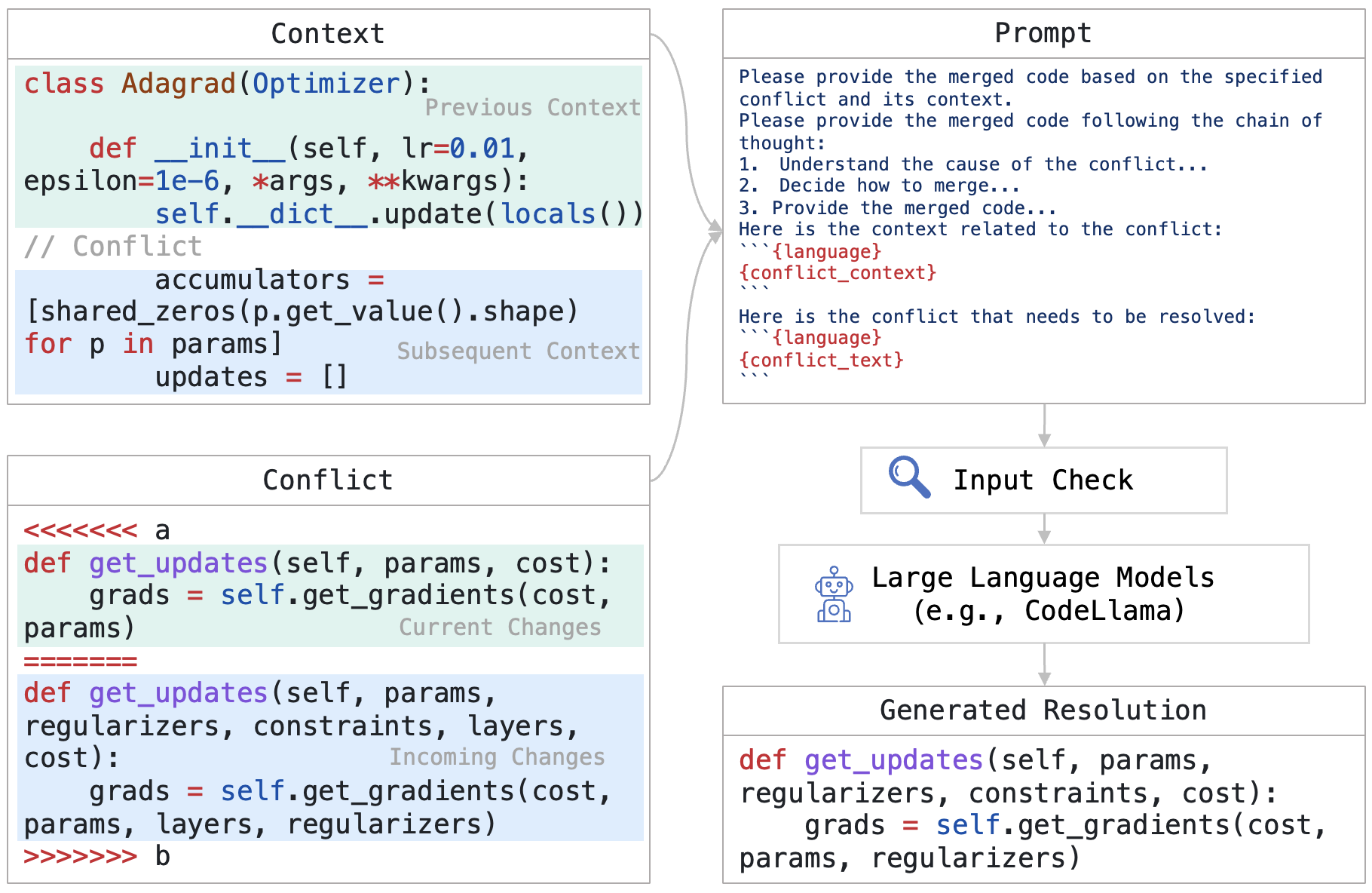}}
	\caption{\textmd{The process of generating resolutions by LLMs.}}
	\label{fg:llm_process}
\end{figure}
In this section, we present how to obtain merge conflict resolution solutions using LLMs, as depicted in ~\autoref{fg:llm_process}. Here, we describe the specific steps as follows: (1) Step 1: Acquire the merge conflict and its context. Merge conflict refers to the current improvements and incoming changes. We select the previous and subsequent text at the conflict location as the context. (2) Step 2: Construct the prompt. We employ the thought-chain method to create the prompt, with the specific content provided in the ~\autoref{sec:prompt}. To avoid outputting redundant context, we offer an example for LLMs to reference. 
(3) Step 3: Check the input. Since LLMs can only support a limited context length, we utilize the corresponding tokenizer to process the prompt. If the number of tokens surpasses the LLMs' maximum input length, revert to Step 1 and reduce the context's lines. (4) Step 4: Query LLMs and obtain the resolution. We input the prompt to LLMs to receive answers, and extract the relevant code block as the final generated resolution, ensuring a logically coherent and standardized format.

\section{Graded Conflicts Dataset}

\begin{figure}[h]
\centerline{\includegraphics[width=\linewidth]{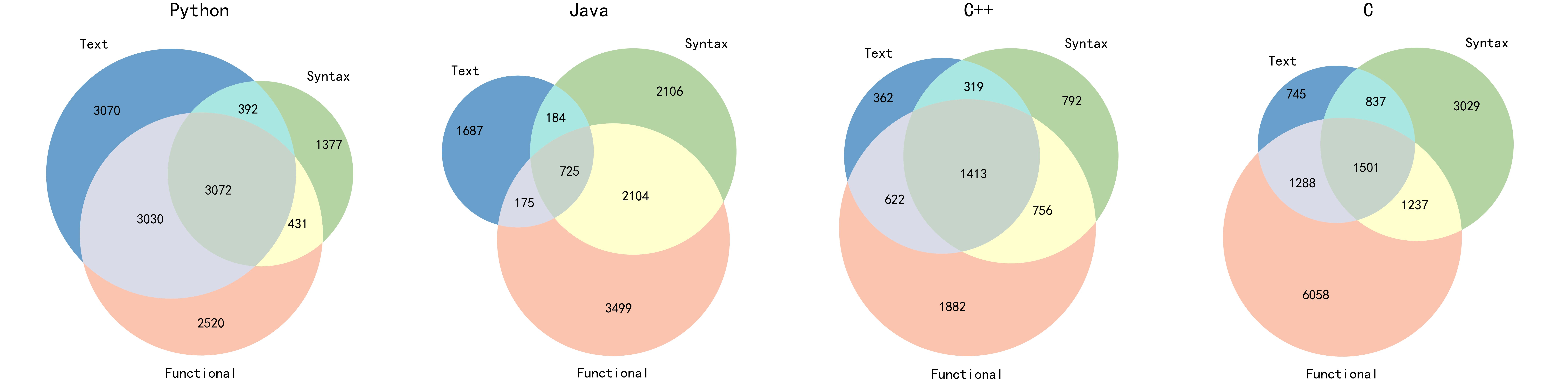}}
	\caption{\textmd{Venn diagrams for conflict types.}}
	\label{fg:overlap}
\end{figure}

\begin{figure}[h]
\centerline{\includegraphics[width=1\linewidth]{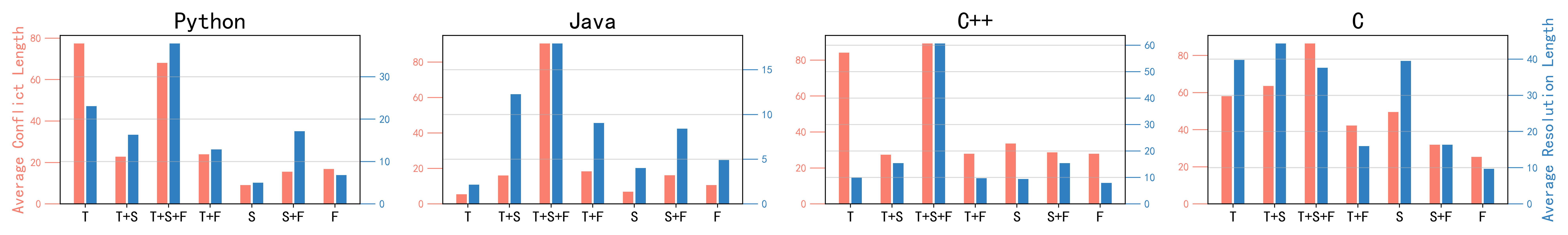}}
	\caption{\textmd{Average length of different types of conflicts. \textbf{T} for Text, \textbf{S} for Syntax, \textbf{F} for Functional.}}
	\label{fg:bar_conflict_length}
\end{figure}

We target on the well-known open-source projects in Github.
Our target selection criteria is that the project should contain 10K+ lines of code and historical commits on main branch.
This is because a larger code base and a more active development history can help generate conflicts with multiple complexities. 
Upon generating the datasets, we first traverse the historical commits of each project and reproduce each merge process through Git merge.
When the merge fails, we record the conflicting file, namely \merged, and extract \va, \vb, \base and \resolved from the commit graph.
Finally, we utilize the coloring and classifier introduced in \S \ref{sec:pipeline} to classify all the collected cases.
%
% We have released two versions of the conflict resolution dataset, named \textsc{ConGra-Full} and \textsc{ConGra-Tiny}. 
%
% We have released the dataset used in \sys benchmarking.
%
% The core difference is that in the \textsc{ConGra-Full} dataset, each file may contain multiple conflicts, while in the \textsc{ConGra-Tiny} dataset, each file is guaranteed to contain only one conflict. 
%
In summary, we collected data from 34 open-source projects on GitHub: 14 written in C/C++, 11 in Java, and 9 in Python.
% %
% For \textsc{ConGra-Full}, we gathered 23,334 conflict files encompassing 44,948 conflict scenarios classified in 7 complexity types as shown in ~\autoref{fg:overlap}.
%
We gathered 23,334 conflict files encompassing 44,948 conflict scenarios classified in 7 complexity types as shown in ~\autoref{fg:overlap}.
The statistics on the length (in lines of code) of all conflicts and their corresponding manual resolutions are presented in ~\autoref{fg:bar_conflict_length}.
More visualizations of \sys are shown in ~\autoref{apd:conflict_num}.
% For \textsc{ConGra-Tiny}, we collected 14,800 conflict files, each containing a single conflict scenario.

% \input{tables/stat_dataset}

% \begin{figure}[h]
% \centerline{\includegraphics[width=0.8\linewidth]{images/conflict_num_box.png}}
% 	\caption{\textmd{Boxplot of the number of conflicts per file.}}
% 	\label{fg:box_conflict_num}
% \end{figure}

\section{Benchmarking for Auto-Conflict Resolutoin}
\subsection{Benchmakring Settings}
\label{sc:setting}

\textbf{Evaluation Metric.}
We choose the accuracy and precision of generated resolutions as the key metrics in our evaluation.
Accuracy is the percentage of generated resolutions that match the ground truth to the total number of conflicts.
Precision refers to the percentage of generated resolutions that match ground truth to all generated resolutions.
We use the combination of normalized edit distance, winnowing, and cosine semantic similarity (\cite{edit-distance, winnowing, cosine-sim}) as the code-matching standard (i.e., ES, WS, and SS).
For a generated resolution, \sys regards the resolution matches the ground truth when at least one of the above values greater than 80\%.

\textbf{Benchmarked Models.}
As depicted in ~\autoref{fg:bar_conflict_length}, \sys exhibits a substantial number of conflict instances, characterized by their contextual information. Consequently, we opt for LLMs that are capable of accommodating long contexts as our benchmarked models. These language models, supporting over 8K tokens, comprise three general language models (\textit{Llama3-8B}~\footnote{\url{https://llama.meta.com/llama3/}}, \textit{DeepSeek-V2}\footnote{We are using their official API service, which supports a context of 32K tokens instead of 128K.}(~\cite{deepseekai2024deepseekv2}), \textit{GLM-3-turbo}~\footnote{\url{https://open.bigmodel.cn/dev/api\#glm-3-turbo}}) and three code language models (\textit{CodeLlama-7B}(~\cite{codellama}), \textit{CodeLlama-34B}(~\cite{codellama}), \textit{DeepSeek-Coder}(~\cite{guo2024deepseekcoder})).

\textbf{Experiment Settings.}
% \label{sc:experiment-setting}
We set the temperature coefficient uniformly at 0.7 to fully ensure the creativity of LLMs in the experiments. For each conflict, we take at most the previous 100 lines and the following 100 lines of text as context, and decrement line by line until the number of tokens is lower than the maximum context length supported by LLMs. In addition, for each conflict that does not correctly obtain the model output, we will repeat the experiment up to 10 times, otherwise it will be regarded as an unprocessable case. For open-source models, we use vLLM(~\cite{vllm} as the backend for model deployment and conduct experiments based on 6 $\times$ NVIDIA A800 80GB GPUs.
% Model Name, Context Length, Param, 
% Considering the typically long length of code files, we choose large language models that support a context length of over 8K as our benchmark models. Our selection of general large language models includes GPT-3.5-Turbo(\texttt{gpt-3.5-turbo-0125},16K), GPT-4(\texttt{gpt-4-0613},8K), GPT-4-Turbo(\texttt{gpt-4-turbo-2024-04-09},128K), LLama3, Deepseek-V2, as well as Deepseek-coder and CodeLLama, which are specifically optimized for code datasets.

% As shown in the ~\autoref{fg:bar_conflict_length},  the conflict in the \sys dataset are 

\subsection{Benchmark Results and Analysis}
%Orange circle (\nag{!}) indicates using non-autoregressive generation. Green circle (\tsc{!}) indicates using both non-autoregressive generation and the proposed two-stage classifier. 

\begin{table}[ht]
\centering
\caption{Benchmark Result on Python and Java. }
\resizebox{\linewidth}{!}{
\begin{tabular}{c|c|ccccc|ccccc}
\hline
                        &                                                                             & \multicolumn{5}{c|}{\textsc{Python}}                                                                                                                                & \multicolumn{5}{c}{\textsc{Java}}                   \\ \cline{3-12} 
\multirow{-2}{*}{Model} & \multirow{-2}{*}{\begin{tabular}[c]{@{}c@{}}Context \\ Length\end{tabular}} & Accuracy                     & Precision                    & ES                          & WS                          & SS                          & Accuracy & Precision & ES   & WS   & SS   \\ \hline
LLama3-8B               & 8K                                                                          & \textbf{75.82}                        & \textbf{77.45}                        & \textbf{0.71}                        & 0.36                        & 0.66                        & 82.93    & 83.00      & \textbf{0.75} & 0.42 & 0.67 \\
DeepSeek-V2             & 32K                                                                         & {75.07} & {75.53} & {0.67} & \textbf{0.53} & {\textbf{0.83}} & \textbf{84.38}    & \textbf{84.40}      & 0.74 & \textbf{0.61} & \textbf{0.84} \\
GLM-3-turbo             & 128K                                                                        & 57.05                        & 57.31                        & 0.53                        & 0.33                        & 0.70                     & 62.52    & 64.75     & 0.58 & 0.38 & 0.74 \\
CodeLlama-7B            & 16K                                                                         & 50.68                        & 59.92                        & 0.55                        & 0.41                        & 0.76                        & 73.61    & 73.66     & 0.67 & 0.51 & 0.79 \\
CodeLlama-34B           & 16K                                                                         & 61.47                        & 62.34                        & 0.61                        & 0.30                     & 0.63                        & 70.82    & 71.03     & 0.66 & 0.40  & 0.70  \\
DeepSeek-Coder          & 16K                                                                         & 56.49                        & 57.31                        & 0.55                        & 0.41                        & 0.76                        & 74.52    & 74.6      & 0.67 & 0.53 & 0.82 \\ \hline
\end{tabular}
}
\label{tb:benchmark_python_java}
\end{table}

\begin{table}[ht]
\centering
\caption{Benchmark Result on C and C++. }
\resizebox{\linewidth}{!}{
\begin{tabular}{c|c|ccccc|ccccc}
\hline
\multirow{2}{*}{Model} & \multirow{2}{*}{\begin{tabular}[c]{@{}c@{}}Context \\ Length\end{tabular}} & \multicolumn{5}{c|}{\textsc{C}}                    & \multicolumn{5}{c}{\textsc{C++}}                   \\ \cline{3-12} 
                       &                                                                            & Accuracy & Precision & ES   & WS   & SS   & Accuracy & Precision & ES   & WS   & SS   \\ \hline
LLama3-8B              & 8K                                                                         & \textbf{72.45}    & \textbf{73.11}     & \textbf{0.64} & 0.36 & 0.69 & \textbf{78.13}    & \textbf{79.22}     & \textbf{0.71} & 0.38 & 0.70  \\
DeepSeek-V2            & 32K                                                                        & 54.42    & 71.06     & 0.61 & \textbf{0.45} & 0.79 & 70.86    & 77.31     & 0.69 & \textbf{0.53} & \textbf{0.83} \\
GLM-3-turbo            & 128K                                                                       & 58.86    & 60.32     & 0.52 & 0.30 & 0.70 & 64.10     & 64.13     & 0.57 & 0.32 & 0.70 \\
CodeLlama-7B           & 16K                                                                        & 59.09    & 60.55     & 0.52 & 0.35 & 0.73 & 64.28    & 64.9      & 0.58 & 0.38 & 0.75 \\
CodeLlama-34B          & 16K                                                                        & 67.75    & 68.23     & 0.61 & 0.26 & 0.63 & 69.83    & 70.39     & 0.66 & 0.29 & 0.63 \\
DeepSeek-Coder         & 16K                                                                        & 62.36    & 62.72     & 0.54 & 0.38 & \textbf{0.77} & 62.33    & 62.9      & 0.57 & 0.39 & 0.78 \\ \hline
\end{tabular}
}
\label{tb:benchmark_c_cpp}
\end{table}
\subsubsection{Overall Performance}
In this section, we present a comprehensive evaluation of advanced LLMs on our proposed \sys dataset. The experimental results can be found in Tables ~\ref{tb:benchmark_python_java} and ~\ref{tb:benchmark_c_cpp}. Upon careful comparison, we draw the following observations: (1) LLMs with longer context support do not always yield optimal results. GLM-3-Turbo, which supports a 128K context, is generally outperformed by LLama3-8B, which only supports an 8K context, with the exception of semantic similarity. For instance, in Python and Java, the Precision metric for GLM-3-Turbo is nearly 20\% lower than that of LLama3-8B. We speculate that this may result from LLMs with ultra-long context support not being sufficiently trained on merge conflict datasets, hindering their ability to effectively extract valuable information from the context to suggest merge resolutions. (2) Code LLMs do not appear to demonstrate a distinct advantage. Notably, LLama3-8B and DeepSeek-V2 surpass Code LLMs in Precision across all four languages. We attribute this to two factors: (i) general models are in fact trained on extensive code repositories, so their code comprehension capabilities are not significantly inferior to Code LLMs. (ii) Elements within conflicts such as comments and variable names contribute a plethora of semantic information, extending beyond mere code understanding, which presents a challenge for Code LLMs. (3) Both LLama3-8B and DeepSeek-V2 prove to be well-suited for automatic conflict resolution.

% \subsection{Benchmark Results and Analysis}
\begin{figure}[!h]
\centerline{\includegraphics[width=1\linewidth]{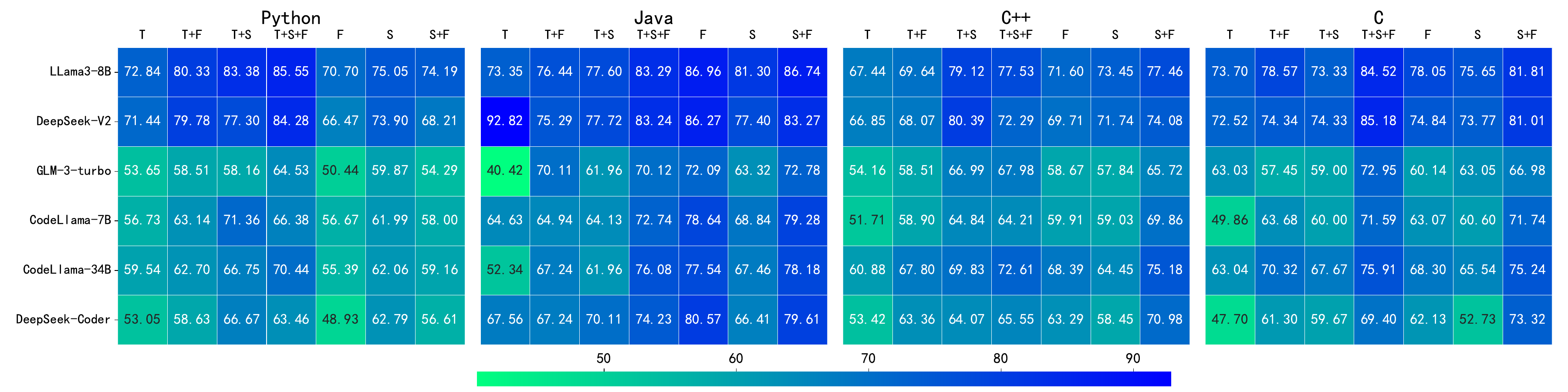}}
	\caption{\textmd{Heatmap on precision of LLMs and types. \textbf{T} for Text, \textbf{S} for Syntax, \textbf{F} for Functional.}}
	\label{fg:precision_heatmap}
\end{figure}

\subsubsection{The Impact of Conflict Type}
% \liangcai{Here, we will describe the result with different conflict types}

Furthermore, we investigate the performance of LLMs under different conflict types. Based on Section ~\ref{sec:classifier}, we consider seven conflict types, namely Text, Syntax, Functional conflicts, and their combinations. The results are visualized in Figure ~\ref{fg:precision_heatmap}. In Figure ~\ref{fg:precision_heatmap}, the closer the result is to \textcolor{blue}{blue}, the better the model performs; the closer it is to \textcolor[HTML]{5dcb9c}{green}, the worse the model performs. % We make the following finding. 

For LLMs, a simpler conflict does not necessarily mean it is easier to handle. This is actually a counterintuitive phenomenon. Specifically, for most models and languages (e.g., Python, C/C++), LLMs exhibit better performance on the most complex conflicts (i.e., F+S+T) and poorer performance on the simplest conflicts (i.e., F, S, and T). In the context of text conflicts, the performance of LLMs on different samples can be ranked as follows: T+F+S $>$ T+F $\approx$ T+S $>$ T. We believe that this occurs because, for LLMs, the simpler the samples within the conflict area (such as type T), the less effective guidance information the conflict can provide, causing LLMs to produce conservative answers. For example, when the conflict involves a comment, LLMs tend to preserve as much information as possible from both branches a and b. In contrast, the more complex the conflict area, the clearer the direction for conflict resolution. In summary, LLMs need to extract more valuable guidance information from the conflict area itself to improve their performance. 
Notice that in \autoref{fg:precision_heatmap}, the precision of F in Java is relatively higher. 
This is due to Java's encapsulation. 
To hide "sensitive" member variables of a class from users, these variables are often declared as private, and the class must provide public \textit{get} and \textit{set} methods to access and update these values. 
Consequently, these getter and setter methods are simple in terms of functionality. 
Since most of the F conflicts in the Java dataset arise from the rewriting of these encapsulated methods, the LLMs can easily infer the resolutions, resulting in high precision.

\subsection{The Role of Context}

\begin{figure}[!h]
\centerline{\includegraphics[width=1\linewidth]{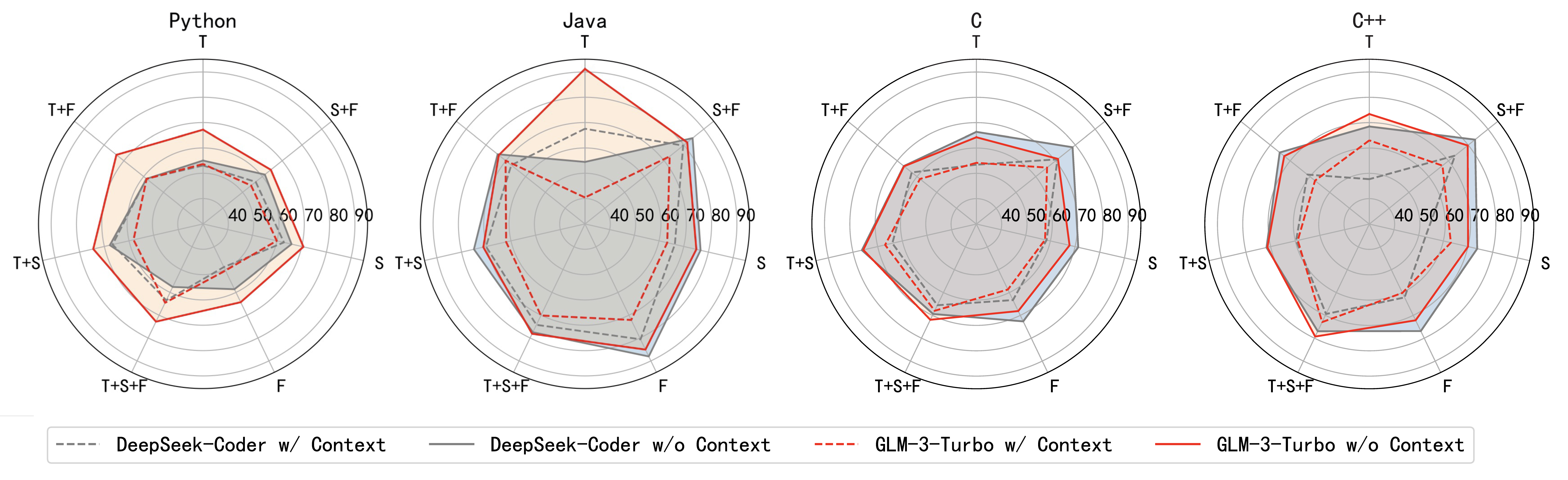}}
	\caption{The impact of contextual information on Precesion.}
	\label{fg:context_exp}
\end{figure}
In the vast majority of cases, LLMs without conflict context information significantly outperform those with context information. As observed in Figure ~\ref{fg:context_exp}, the red and gray solid circles almost entirely encompass their corresponding dashed circles, with only a few exceptions. We analyze this phenomenon from two perspectives: (1) the context understanding capability of large language models. Although existing LLMs have undergone extensive training on code data, this does not necessarily imply that they can effectively extract useful information from the rich context for automatic conflict resolution. (2) The choice of context. In our experiments, we provided the most basic context information. However, it is possible that the adjacent context does not supply crucial information for resolving conflicts and instead introduces a significant amount of noise. As a result, selecting the appropriate context for automatic conflict resolution may be a valuable research direction to explore.

% \subsection{Cost Analysis}

% \input{sections/related_work}
\section{Discussion}
\label{sc:discussion}

\textbf{Limitation. }
The conflict classification strategy in \S \ref{sec:pipeline} ensures correct classification but may fail to identify the exact category of a conflict due to advanced language usage.
For example, the use of template class definitions in C++ (\cite{vandevoorde2002c++}) prevents tree\_sitter (\cite{tree-sitter}) from capturing the type used as the template argument during instantiation. 
Consequently, a conflict that should be classified as "syntax" may be regarded as "non-syntax."
We will continue to refine \sys's classification in future work.

\textbf{Societal Impacts. }
Our work alerts the software engineering and machine learning communities to the problem of code merging, without obvious negative societal impact.

\section{Conclusion}

We propose \sys, a complexity-graded conflict benchmarking system.
\sys implements a highly efficient and accurate conflict classification algorithm to construct a complexity-graded conflict dataset, which is used to evaluate the performance of merging tools under various conflict scenarios. 
\sys utilizes three code matching metrics of different granularities and combines them to calculate the accuracy and precision of auto-generated resolutions. 
We evaluate six LLMs on \sys, and the results show that LLMs with longer context support often perform worse than those with shorter context support, and general LLMs outperform specialized code LLMs in precision.

% \begin{ack}
% Use unnumbered first level headings for the acknowledgments. All acknowledgments
% go at the end of the paper before the list of references. Moreover, you are required to declare
% funding (financial activities supporting the submitted work) and competing interests (related financial activities outside the submitted work).
% More information about this disclosure can be found at: \url{https://neurips.cc/Conferences/2024/PaperInformation/FundingDisclosure}.
% You can use the \texttt{ack} environment provided in the style file. As opposed to the main NeurIPS track, acknowledgements do not need to be hidden.
% \end{ack}

\newpage
\bibliographystyle{plainnat}
\bibliography{reference}
% \medskip

%%%%%%%%%%%%%%%%%%%%%%%%%%%%%%%%%%%%%%%%%%%%%%%%%%%%%%%%%%%%

\appendix
\newpage
\section{Prompt}
\label{sec:prompt}
\begin{figure}[h]
\centerline{\includegraphics[width=\linewidth]{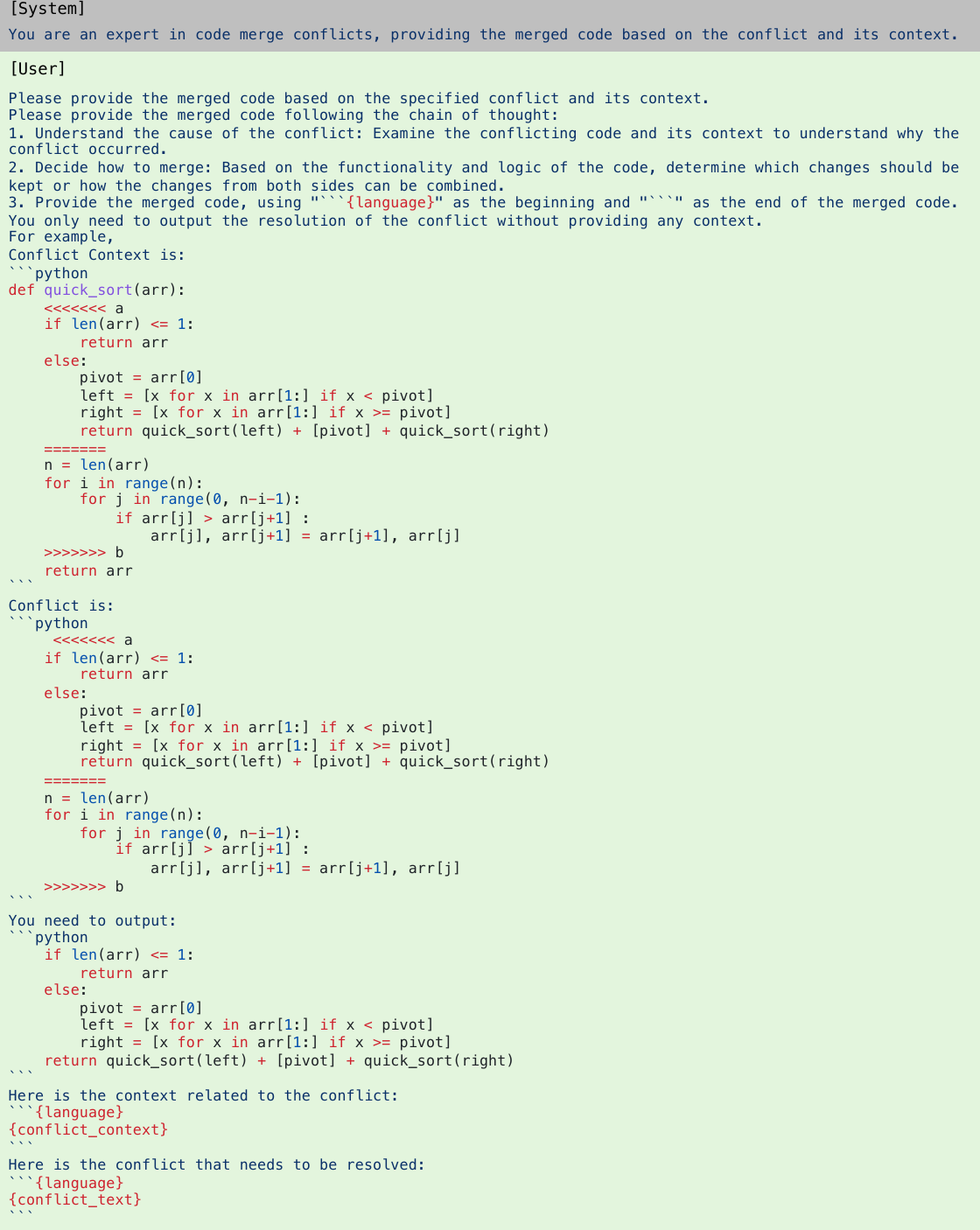}}
	\caption{Prompt}
	\label{fg:prompt}
\end{figure}

\section{Demo Case}

\subsection{Conflict}

\begin{lstlisting}[label={conflict-demo-py-text}, language=diff, caption={\textmd{Demo 1: text conflict}}]
<<<<<<< a
        padding: Int, or tuple of 3 ints, or tuple of 3 tuples of 2 ints.
            - If int: the same symmetric padding is applied to depth, height,
              and width.
            - If tuple of 3 ints: interpreted as three different symmetric
              padding values for depth, height, and width:
              `(symmetric_dim1_pad, symmetric_dim2_pad, symmetric_dim3_pad)`.
            - If tuple of 3 tuples of 2 ints: interpreted as
              `((left_dim1_pad, right_dim1_pad), (left_dim2_pad,
              right_dim2_pad), (left_dim3_pad, right_dim3_pad))`.
        data_format: A string, one of `"channels_last"` (default) or
            `"channels_first"`. The ordering of the dimensions in the inputs.
            `"channels_last"` corresponds to inputs with shape
            `(batch_size, spatial_dim1, spatial_dim2, spatial_dim3, channels)`
            while `"channels_first"` corresponds to inputs with shape
            `(batch_size, channels, spatial_dim1, spatial_dim2, spatial_dim3)`.
            When unspecified, uses `image_data_format` value found in your Keras
            config file at `~/.keras/keras.json` (if exists). Defaults to
            `"channels_last"`.
=======
      padding: Int, or tuple of 3 ints, or tuple of 3 tuples of 2 ints.
        - If int: the same symmetric padding
          is applied to height and width.
        - If tuple of 3 ints:
          interpreted as two different
          symmetric padding values for height and width:
          `(symmetric_dim1_pad, symmetric_dim2_pad, symmetric_dim3_pad)`.
        - If tuple of 3 tuples of 2 ints:
          interpreted as
          `((left_dim1_pad, right_dim1_pad), (left_dim2_pad,
            right_dim2_pad), (left_dim3_pad, right_dim3_pad))`
      data_format: A string,
        one of `channels_last` (default) or `channels_first`.
        The ordering of the dimensions in the inputs.
        `channels_last` corresponds to inputs with shape
        `(batch_size, spatial_dim1, spatial_dim2, spatial_dim3, channels)`
        while `channels_first` corresponds to inputs with shape
        `(batch_size, channels, spatial_dim1, spatial_dim2, spatial_dim3)`.
        It defaults to the `image_data_format` value found in your
        Keras config file at `~/.keras/keras.json`.
        If you never set it, then it will be "channels_last".
>>>>>>> b
\end{lstlisting}

\begin{lstlisting}[label={conflict-demo-py-text}, language=diff, caption={\textmd{Demo 2: text and functional conflict}}]
<<<<<<< a
    constraints = _process_dynamic_shapes(mod, args, kwargs, dynamic_shapes) or []

    kwargs = kwargs or {}
=======
    if constraints is not None:
        log_export_usage(event="export.private_api", flags={"constraints"})
        warnings.warn(
            "Using `constraints` to specify dynamic shapes for export is DEPRECATED "
            "and will not be supported in the future. "
            "Please use `dynamic_shapes` instead (see docs on `torch.export.export`).",
            DeprecationWarning,
            stacklevel=2,
        )
    else:
        constraints = _process_dynamic_shapes(f, args, kwargs, dynamic_shapes) or []
>>>>>>> b
\end{lstlisting}

\begin{lstlisting}[label={conflict-demo-py-text}, language=diff, caption={\textmd{Demo 3: syntax conflict}}]
<<<<<<< a
import torch.utils._pytree as pytree
=======
>>>>>>> b
from torch._decomp import register_decomposition
\end{lstlisting}

\begin{lstlisting}[label={conflict-demo-py-text}, language=diff, caption={\textmd{Demo 4: text, syntax, and functional conflict}}]
<<<<<<< a
            log.info("converting frame raised error, suppressing error")
=======

            # Suppress the error.  NB: It's very important to do the
            # suppression logging HERE, where the actual suppression
            # happens. Previously it was somewhere else and so it was
            # possible to accidentally not log at all.
            record_filename = getattr(e, "record_filename", None)
            code = frame.f_code
            if config.is_fbcode():
                from torch._dynamo.fb.logging import (  # type: ignore[import]
                    log_dynamo_suppress_errors,
                )

                error_msg = format_error_msg_verbose(e, code, record_filename, frame)
                log_dynamo_suppress_errors(
                    code.co_name, code.co_filename, code.co_firstlineno, error_msg
                )
            else:
                error_msg = format_error_msg(e, code, record_filename, frame)

            if soft_fail:
                log.info(error_msg, exc_info=True)
            else:
                log.warning(error_msg, exc_info=True)
>>>>>>> b
        return None
\end{lstlisting}

\subsection{Ground Truth}

\begin{lstlisting}[label={groundtruth-demo-py-text}, breaklines, language={}, caption={\textmd{Ground truth of demo 1}}]
    Args:
        padding: Int, or tuple of 3 ints, or tuple of 3 tuples of 2 ints.
            - If int: the same symmetric padding is applied to depth, height,
              and width.
            - If tuple of 3 ints: interpreted as three different symmetric
              padding values for depth, height, and width:
              `(symmetric_dim1_pad, symmetric_dim2_pad, symmetric_dim3_pad)`.
            - If tuple of 3 tuples of 2 ints: interpreted as
              `((left_dim1_pad, right_dim1_pad), (left_dim2_pad,
              right_dim2_pad), (left_dim3_pad, right_dim3_pad))`.
        data_format: A string, one of `"channels_last"` (default) or
            `"channels_first"`. The ordering of the dimensions in the inputs.
            `"channels_last"` corresponds to inputs with shape
            `(batch_size, spatial_dim1, spatial_dim2, spatial_dim3, channels)`
            while `"channels_first"` corresponds to inputs with shape
            `(batch_size, channels, spatial_dim1, spatial_dim2, spatial_dim3)`.
            When unspecified, uses `image_data_format` value found in your Keras
            config file at `~/.keras/keras.json` (if exists). Defaults to
            `"channels_last"`.
\end{lstlisting}

\begin{lstlisting}[label={groundtruth-demo-py-text}, breaklines, language={python}, caption={\textmd{Ground truth of demo 2}}]

    kwargs = kwargs or {}
    _process_dynamic_shapes(mod, args, kwargs, dynamic_shapes)  # TODO(avik): remove
\end{lstlisting}

\begin{lstlisting}[label={groundtruth-demo-py-text}, breaklines, language={python}, caption={\textmd{Ground truth of demo 3}}]
import torch._inductor as inductor
import torch.utils._pytree as pytree
from torch import fx
from torch._decomp import register_decomposition
\end{lstlisting}

\begin{lstlisting}[label={groundtruth-demo-py-text}, breaklines, language={python}, caption={\textmd{Ground truth of demo 4}}]
                raise

            # Suppress the error.  NB: It's very important to do the
            # suppression logging HERE, where the actual suppression
            # happens. Previously it was somewhere else and so it was
            # possible to accidentally not log at all.
            record_filename = getattr(e, "record_filename", None)
            code = frame.f_code
            error_msg = format_error_msg(e, code, record_filename, frame)

            if soft_fail:
                log.info(error_msg, exc_info=True)
            else:
                log.warning(error_msg, exc_info=True)
\end{lstlisting}

\subsection{Resolution generated by LLama3-8B}

\begin{lstlisting}[label={llama3-demo-py-text}, breaklines, language={}, caption={\textmd{Demo 1: LLama3-8B resolution}}]
    padding: Int, or tuple of 3 ints, or tuple of 3 tuples of 2 ints.
        - If int: the same symmetric padding is applied to depth, height, and width.
        - If tuple of 3 ints: interpreted as three different symmetric padding values for depth, height, and width:
          `(symmetric_dim1_pad, symmetric_dim2_pad, symmetric_dim3_pad)`.
        - If tuple of 3 tuples of 2 ints: interpreted as
          `((left_dim1_pad, right_dim1_pad), (left_dim2_pad, right_dim2_pad), (left_dim3_pad, right_dim3_pad))`.
    data_format: A string, one of `"channels_last"` (default) or
        `"channels_first"`. The ordering of the dimensions in the inputs.
        `"channels_last"` corresponds to inputs with shape
        `(batch_size, spatial_dim1, spatial_dim2, spatial_dim3, channels)`
        while `"channels_first"` corresponds to inputs with shape
        `(batch_size, channels, spatial_dim1, spatial_dim2, spatial_dim3)`.
        When unspecified, uses `image_data_format` value found in your Keras
        config file at `~/.keras/keras.json` (if exists). Defaults to
        `"channels_last"`.
\end{lstlisting}

\begin{lstlisting}[label={llama3-demo-py-text}, breaklines, language={python}, caption={\textmd{Demo 2: LLama3-8B resolution}}]
    if constraints is not None:
        log_export_usage(event="export.private_api", flags={"constraints"})
        warnings.warn(
            "Using `constraints` to specify dynamic shapes for export is DEPRECATED "
            "and will not be supported in the future. "
            "Please use `dynamic_shapes` instead (see docs on `torch.export.export`).",
            DeprecationWarning,
            stacklevel=2,
        )
    else:
        constraints = _process_dynamic_shapes(f, args, kwargs, dynamic_shapes) or []
    kwargs = kwargs or {}

\end{lstlisting}

\begin{lstlisting}[label={llama3-demo-py-text}, breaklines, language={python}, caption={\textmd{Demo 3: LLama3-8B resolution}}]
from torch.utils._pytree import pytree

\end{lstlisting}

\begin{lstlisting}[label={llama3-demo-py-text}, breaklines, language={python}, caption={\textmd{Demo 4: LLama3-8B resolution}}]
            log.info("converting frame raised error, suppressing error")

\end{lstlisting}

\subsection{Resolution generated by Deepseek-Chat}

\begin{lstlisting}[label={chat-demo-py-text}, language={}, caption={\textmd{Demo 1: Deepseek-Chat resolution}}]
        padding: Int, or tuple of 3 ints, or tuple of 3 tuples of 2 ints.
            - If int: the same symmetric padding is applied to depth, height,
              and width.
            - If tuple of 3 ints: interpreted as three different symmetric
              padding values for depth, height, and width:
              `(symmetric_dim1_pad, symmetric_dim2_pad, symmetric_dim3_pad)`.
            - If tuple of 3 tuples of 2 ints: interpreted as
              `((left_dim1_pad, right_dim1_pad), (left_dim2_pad,
              right_dim2_pad), (left_dim3_pad, right_dim3_pad))`.
        data_format: A string, one of `"channels_last"` (default) or
            `"channels_first"`. The ordering of the dimensions in the inputs.
            `"channels_last"` corresponds to inputs with shape
            `(batch_size, spatial_dim1, spatial_dim2, spatial_dim3, channels)`
            while `"channels_first"` corresponds to inputs with shape
            `(batch_size, channels, spatial_dim1, spatial_dim2, spatial_dim3)`.
            It defaults to the `image_data_format` value found in your
            Keras config file at `~/.keras/keras.json`.
            If you never set it, then it will be "channels_last".
\end{lstlisting}

\begin{lstlisting}[label={chat-demo-py-text}, language={python}, caption={\textmd{Demo 2: Deepseek-Chat resolution}}]
    if constraints is not None:
        log_export_usage(event="export.private_api", flags={"constraints"})
        warnings.warn(
            "Using `constraints` to specify dynamic shapes for export is DEPRECATED "
            "and will not be supported in the future. "
            "Please use `dynamic_shapes` instead (see docs on `torch.export.export`).",
            DeprecationWarning,
            stacklevel=2,
        )
    else:
        constraints = _process_dynamic_shapes(f, args, kwargs, dynamic_shapes) or []

\end{lstlisting}

\begin{lstlisting}[label={chat-demo-py-text}, language={python}, caption={\textmd{Demo 3: Deepseek-Chat resolution}}]
import torch.utils._pytree as pytree
from torch._decomp import register_decomposition

\end{lstlisting}

\begin{lstlisting}[label={chat-demo-py-text}, language={python}, caption={\textmd{Demo 4: Deepseek-Chat resolution}}]
            # Suppress the error.  NB: It's very important to do the
            # suppression logging HERE, where the actual suppression
            # happens. Previously it was somewhere else and so it was
            # possible to accidentally not log at all.
            record_filename = getattr(e, "record_filename", None)
            code = frame.f_code
            if config.is_fbcode():
                from torch._dynamo.fb.logging import (  # type: ignore[import]
                    log_dynamo_suppress_errors,
                )

                error_msg = format_error_msg_verbose(e, code, record_filename, frame)
                log_dynamo_suppress_errors(
                    code.co_name, code.co_filename, code.co_firstlineno, error_msg
                )
            else:
                error_msg = format_error_msg(e, code, record_filename, frame)

            if soft_fail:
                log.info(error_msg, exc_info=True)
            else:
                log.warning(error_msg, exc_info=True)
        return None

\end{lstlisting}

\subsection{Resolution generated by Deepseek-Coder}

\begin{lstlisting}[label={coder-demo-py-text}, language={}, caption={\textmd{Demo 1: Deepseek-Coder resolution}}]
    padding: Int, or tuple of 3 ints, or tuple of 3 tuples of 2 ints.
        - If int: the same symmetric padding is applied to depth, height,
          and width.
        - If tuple of 3 ints: interpreted as three different symmetric
          padding values for depth, height, and width:
          `(symmetric_dim1_pad, symmetric_dim2_pad, symmetric_dim3_pad)`.
        - If tuple of 3 tuples of 2 ints: interpreted as
          `((left_dim1_pad, right_dim1_pad), (left_dim2_pad,
            right_dim2_pad), (left_dim3_pad, right_dim3_pad))`.
    data_format: A string, one of `"channels_last"` (default) or
        `"channels_first"`. The ordering of the dimensions in the inputs.
        `"channels_last"` corresponds to inputs with shape
        `(batch_size, spatial_dim1, spatial_dim2, spatial_dim3, channels)`
        while `"channels_first"` corresponds to inputs with shape
        `(batch_size, channels, spatial_dim1, spatial_dim2, spatial_dim3)`.
        When unspecified, uses `image_data_format` value found in your Keras
        config file at `~/.keras/keras.json` (if exists). Defaults to
        `"channels_last"`.
\end{lstlisting}

\begin{lstlisting}[label={coder-demo-py-text}, language={python}, caption={\textmd{Demo 2: Deepseek-Coder resolution}}]
    constraints = _process_dynamic_shapes(f, args, kwargs, dynamic_shapes) or []

    kwargs = kwargs or {}

\end{lstlisting}

\begin{lstlisting}[label={coder-demo-py-text}, language={python}, caption={\textmd{Demo 3: Deepseek-Coder resolution}}]
import torch.utils._pytree as pytree
from torch._decomp import register_decomposition

\end{lstlisting}

\begin{lstlisting}[label={coder-demo-py-text}, language={python}, caption={\textmd{Demo 4: Deepseek-Coder resolution}}]
            log.info("converting frame raised error, suppressing error")
            record_filename = getattr(e, "record_filename", None)
            code = frame.f_code
            if config.is_fbcode():
                from torch._dynamo.fb.logging import (  # type: ignore[import]
                    log_dynamo_suppress_errors,
                )

                error_msg = format_error_msg_verbose(e, code, record_filename, frame)
                log_dynamo_suppress_errors(
                    code.co_name, code.co_filename, code.co_firstlineno, error_msg
                )
            else:
                error_msg = format_error_msg(e, code, record_filename, frame)

            if soft_fail:
                log.info(error_msg, exc_info=True)
            else:
                log.warning(error_msg, exc_info=True)
            return None

\end{lstlisting}

\section{Git Merge Conflict Example}
\label{apd:conflict-example}
\begin{figure}[!h]
    \centering
    \begin{subfigure}{0.32\textwidth}
        \includegraphics[width=\textwidth]{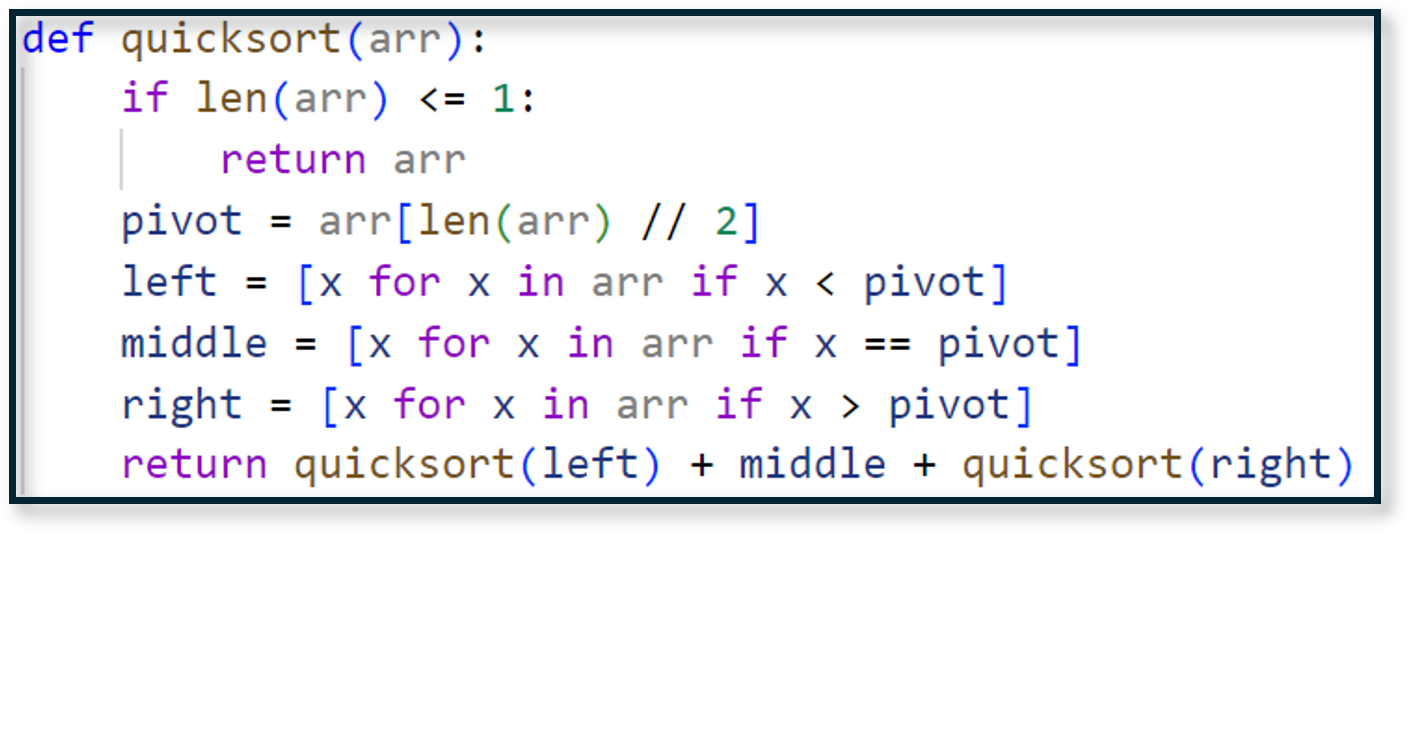}
        \caption{Quick sort from version A.}
        \label{fig:code-merge-conflict-a}
    \end{subfigure}
    \hfill
    \begin{subfigure}{0.32\textwidth}
        \includegraphics[width=\textwidth]{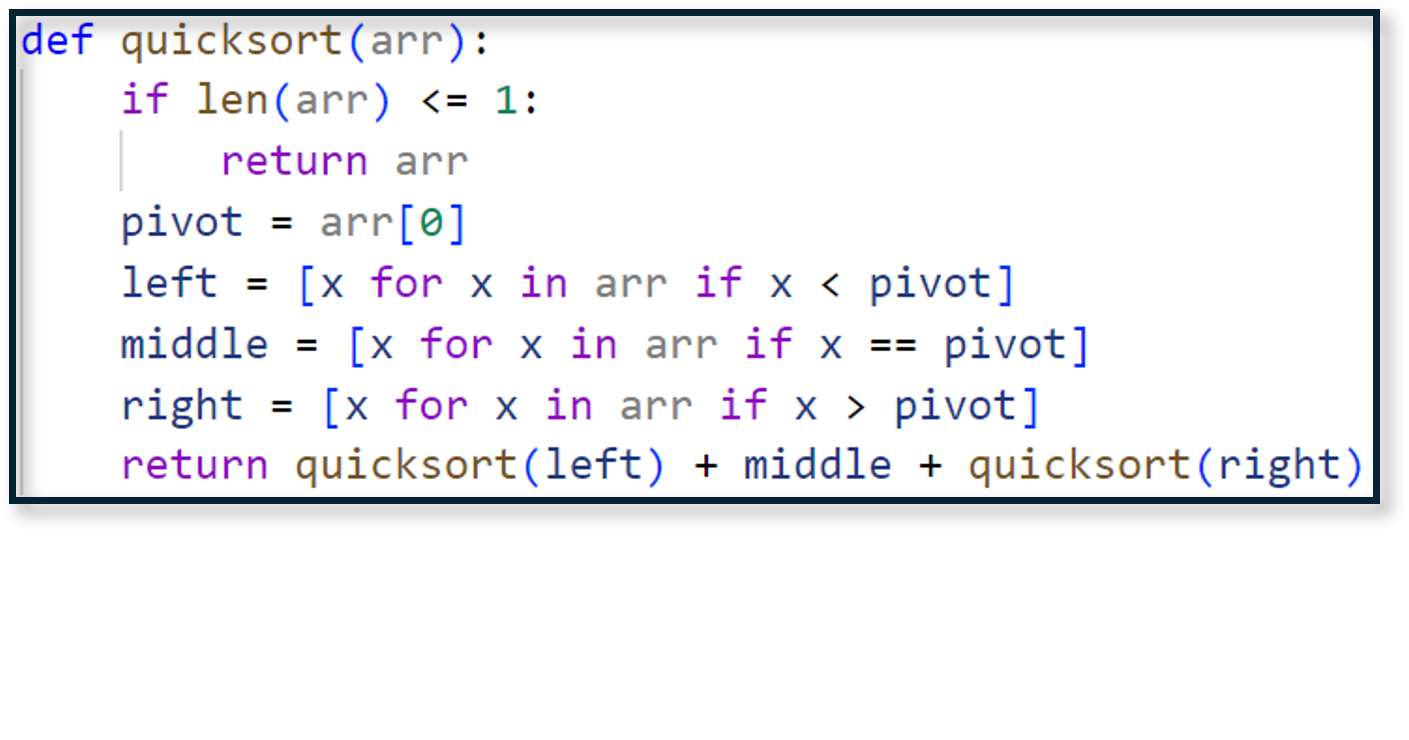}
        \caption{Quick sort from version B.}
        \label{fig:code-merge-conflict-b}
    \end{subfigure}
    \hfill
    \begin{subfigure}{0.32\textwidth}
        \includegraphics[width=\textwidth]{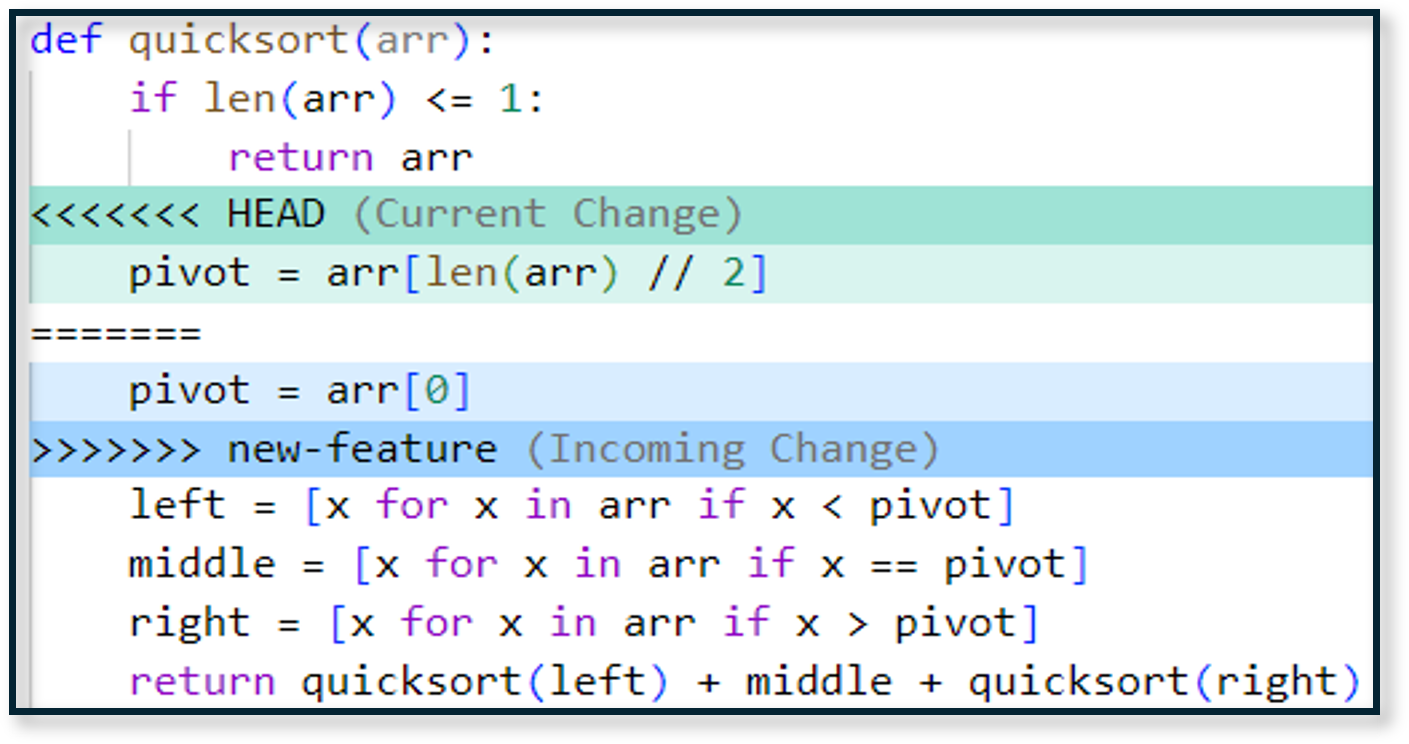}
        \caption{Quick sort with merge conflict.}
        \label{fig:code-merge-conflict-merged}
    \end{subfigure}
    \caption{Quick sort merging example}
    \label{fig:code-merge-conflict}
\end{figure}

~\autoref{fig:code-merge-conflict} shows a code merging scenario with conflict. 
Both version A (\autoref{fig:code-merge-conflict-a}) and version B (~\autoref{fig:code-merge-conflict-b}) implement quick sort, but version A selects the middle element of \textit{arr} as the pivot, while version B selects the first element.
This discrepancy causes Git to encounter an impasse and report a conflict, as depicted in ~\autoref{fig:code-merge-conflict-merged}.

\section{License}
\label{sc:license}

We utilized the source code of 34 open-source projects in this research. These projects and their license information are as follows:
1) Linux Kernel, Android Kernel, Raspberry Pi Kernel, Git, MySQL, ReactOS, JDK, NewPipe, Ansible Cpython, are licensed under GNU General Public License (GPL);
2) Bitcoin, GCC, Jenkins, are licensed under the Massachusetts Institute of Technology (MIT);
3) LLVM, Swift, Tensorflow, AOSP, dbeaver, Ghidra, hadoop, Micronaut, Netty, Sprint-boot, Sprint-framework, Keras, Transformers, are under the Apache License;
4) Mongo is under Server Side Public License;
5) PHP is under PHP License;
6) V8, Django, Pandas, Pytorch, Scrapy, are under BSD-3-Clause License;
7) Eclipse is under Eclipse Public License;
8) Youtube-dl is unlicensed.
We acknowledge the contributions of the open-source community in developing and maintaining these projects.

\section{Code Similarity Metrics}
\label{apd:code-sim-metrics}

\begin{itemize}
    \item \textbf{Edit Similarity (ES)}. Edit distance is a metric used to measure the difference between two strings (\cite{edit-distance}). 
    It is defined as the minimum number of edit operations required to transform one string into another. 
    In \sys, we normalize the edit distance, treating the normalized result as the similarity measure in character level.
    
    \item \textbf{Winnowing Similarity (WS)}. Winnowing is an algorithm used for text similarity detection and fingerprint extraction (\cite{winnowing}). 
    It has been implemented in the Moss code plagiarism detection system (\cite{moss}) to determine code similarity.
    In \sys, we normalize the winnowing result to assess the similarity of the whole conflict code snippets.

    \item \textbf{Semantic Similarity (SS)}.  We use the cosine similarity between the generated resolution and the actual resolution as the semantic similarity. In \sys, in order to effectively model the semantic information of the resolution, we use \texttt{BCEmbedding}~\cite{BCEEMbeddings} as the embedding model to obtain the representation of each resolution.
    %Cosine semantic similarity measures the similarity between two text vectors by calculating the cosine of the angle between them (\cite{cosine-sim}). 
    %
    % It represents text data in a high-dimensional space and evaluates how similar the content is based on the direction of the vectors, regardless of their magnitude. 
    %
    % In \sys, we use cosine similarity to compare the code semantic similarity of conflict regions.
\end{itemize}

\section{Visualization of the number of conflicts in \sys}\label{apd:conflict_num}
\begin{figure}[h]
\centerline{\includegraphics[width=\linewidth]{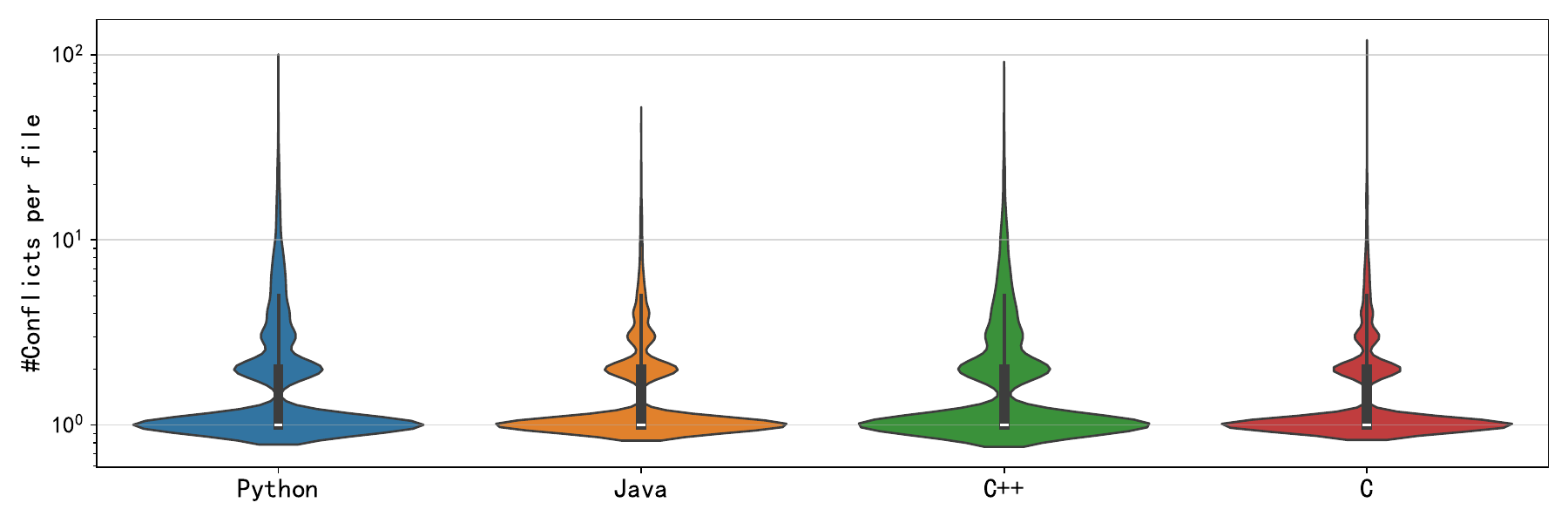}}
	\caption{\textmd{Violin plots of the number of conflicts per file.}}
	\label{fg:box_conflict_num}
\end{figure}

\begin{figure}
    \centering
    \begin{subfigure}[b]{\textwidth}
        \includegraphics[width=0.95\textwidth]{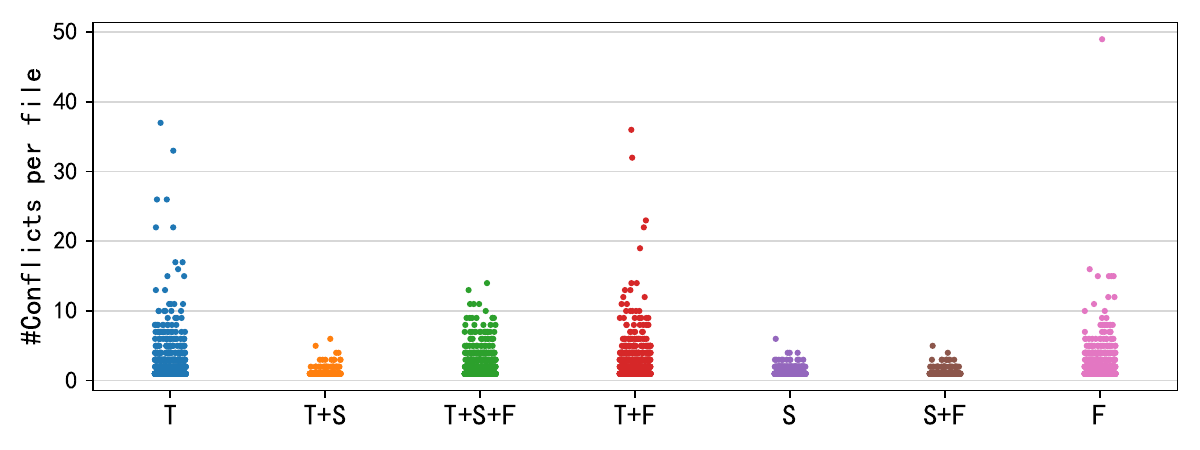}
        \caption{Python}
        \label{fig:sub1}
    \end{subfigure}
    \hfill
    \begin{subfigure}[b]{\textwidth}
        \includegraphics[width=0.95\textwidth]{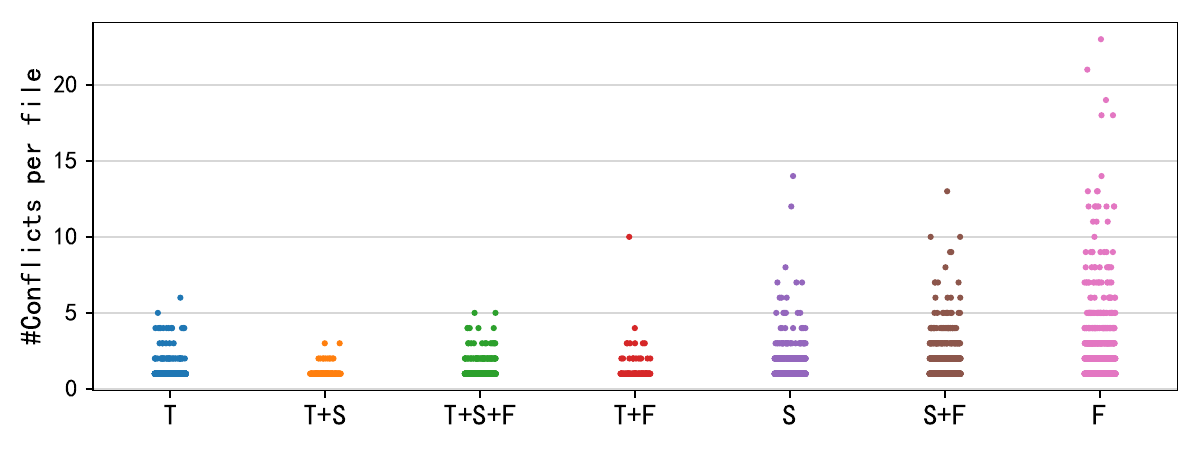}
        \caption{Java}
        \label{fig:sub2}
    \end{subfigure}
    \hfill
    \begin{subfigure}[b]{\textwidth}
        \includegraphics[width=0.95\textwidth]{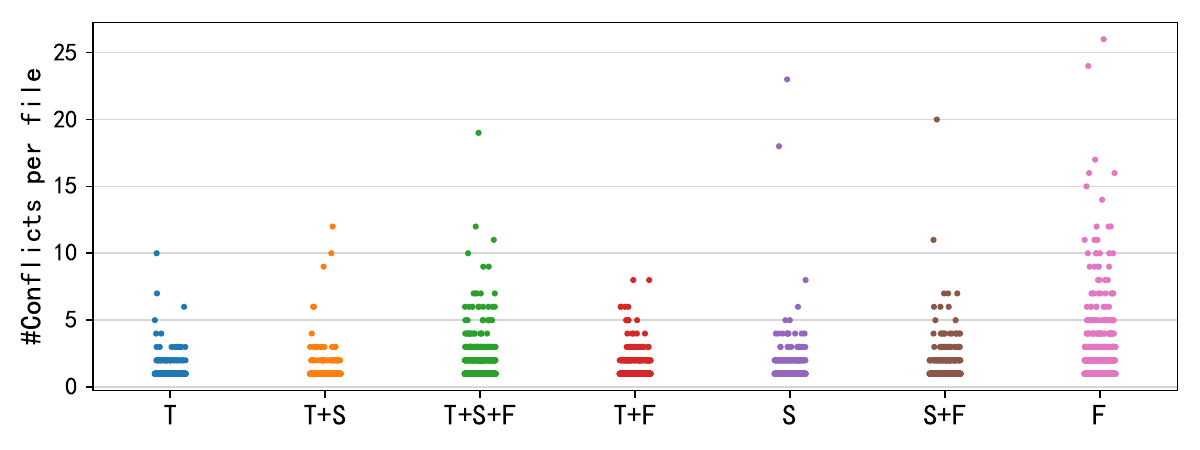}
        \caption{C++}
        \label{fig:sub3}
    \end{subfigure}
    \hfill
    \begin{subfigure}[b]{\textwidth}
        \includegraphics[width=0.95\textwidth]{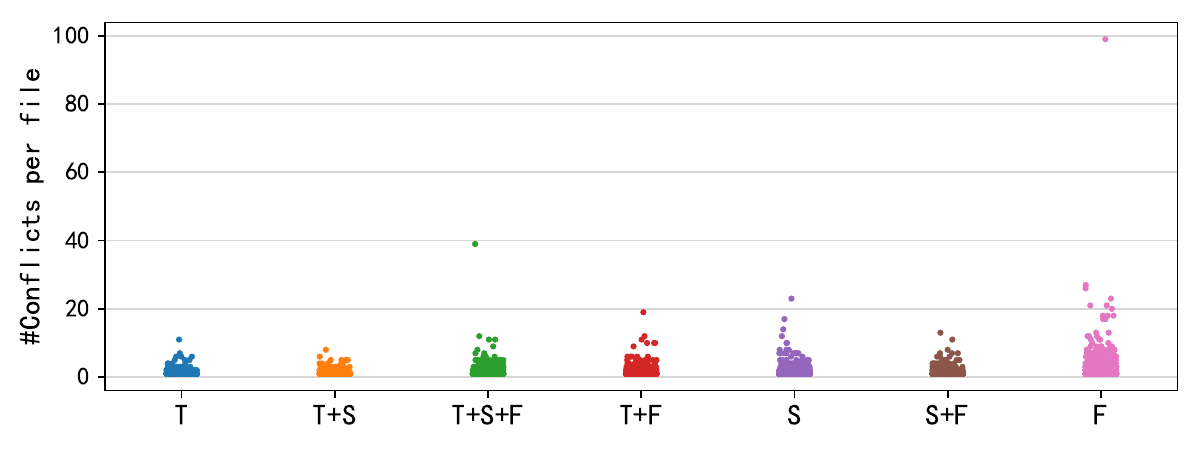}
        \caption{C}
        \label{fig:sub4}
    \end{subfigure}
    \caption{Strip plots of number of conflicts per file.}
    \label{fg:strip_conflict_num}
\end{figure}

Here we visualize the distribution characteristics of the number of conflicts contained in each file in four languages in ~\autoref{fg:box_conflict_num}. Overall, most files contain only one or two conflicts, and a very small number of files contain more than 10 conflicts. We also visualize the number of different types of conflicts for each language in ~\autoref{fg:strip_conflict_num}.

\end{document}